\documentclass[letter,preprint,superscriptaddress,nofootinbib,longbibliography]{revtex4-1}
\usepackage[dvipdfmx]{graphicx}
\usepackage{bm}
\usepackage{amssymb}
\usepackage{amsmath}
\allowdisplaybreaks
\usepackage{color}
\usepackage{cancel}
\usepackage{comment}
\usepackage{here}
\usepackage{hyperref}
\usepackage{subfigure}
\usepackage{multirow}
\begin{document} 

\title{Oblique corrections in general dark $U(1)$ models}
\author{Cheng-Wei Chiang}
\email{chengwei@phys.ntu.edu.tw}
\affiliation{Department of Physics and Center for Theoretical Physics, National Taiwan University, Taipei, 10617, Taiwan}
\affiliation{Physics Division, National Center for Theoretical Sciences, Taipei, 10617, Taiwan}

\author{Kazuki Enomoto}
\email{k\_enomoto@phys.ntu.edu.tw}
\affiliation{Department of Physics and Center for Theoretical Physics, National Taiwan University, Taipei, 10617, Taiwan}

\begin{abstract}
We investigate the impact of dark Abelian gauge bosons on the electroweak precision measurements at the one-loop level.  
The dark gauge boson couples to the standard model fermions generally via two kinds of mixing with the electroweak gauge bosons: the kinetic mixing and the mass mixing.   
We solve the Schwinger--Dyson equation for the gauge boson propagators and derive a renormalization scheme-independent representation of the scattering amplitudes for four-fermion processes, including the full oblique corrections. 
We define the running parameters at the one-loop level and show that the leading new physics effects, including the mixing, in the electroweak precision observables can be described by the oblique parameters $S$, $T$, and $U$ as in the standard electroweak gauge theory when the new physics scale is sufficiently high and the dark gauge boson mass lies away from the $Z$ pole. We consider the dark doublet scalar boson as an example and numerically show that a novel one-loop effect can drastically change the parameter region allowed by the electroweak precision tests.
\end{abstract}

\maketitle

%%%%%%%%%%%%%%%%%%%%%%%%%%%%%%%%%%%%%%%%%%%%%%%%%%
\section{Introduction}
%%%%%%%%%%%%%%%%%%%%%%%%%%%%%%%%%%%%%%%%%%%%%%%%%%

There is mounting evidence for the quest of physics beyond the standard model (BSM) from both theoretical problems in the standard model (SM) and observed BSM phenomena such as neutrino oscillations, the existence of dark matter, and the baryon asymmetry of the Universe~\cite{PDG}. 
Various new physics models have been proposed to explain such issues.
A thorough investigation of the phenomenology associated with such models in current and future experiments has been one of the most active researched areas in particle physics.

New physics models predict not only direct signals of new particles but also indirect evidence as deviations from the SM expectations in the precision measurements of specific observables.  
As an example, the current data of electroweak precision observables (EWPOs) impose a severe constraint on various BSM models and thus play a significant role in guiding the search for new physics~\cite{PDG, ALEPH:2005ab, deBlas:2016ojx}. 
A future collider experiment for precision measurements at the $Z$ pole has been proposed~\cite{Erler:2000jg}.

Assuming that new particles do not directly couple to the SM fermions, their leading effects appear in oblique corrections, {\it i.e.}, loop corrections to gauge boson propagators, in four-fermion processes~\cite{Kennedy:1988sn, Hagiwara:1994pw}. 
In the standard electroweak (EW) gauge sector $SU(2)_L\times U(1)_Y$, the oblique corrections can be described by three oblique parameters~\cite{Kennedy:1988sn, Peskin:1990zt, Kennedy:1990ib, Marciano:1990dp, Altarelli:1990zd}, which are usually parametrized by $S$, $T$, and $U$~\cite{Peskin:1990zt}, when the scale of new physics is much higher than the $Z$ boson mass. 
Although there are some limitations on new physics models that can utilize these parameters, they enable a model-independent analysis to examine the EWPO constraint instead of the global fit of the full BSM models. 
Extensions of the oblique parameters have been discussed in Refs.~\cite{Maksymyk:1993zm, Burgess:1993mg, Drauksas:2023isd, Long:1999bny, Cui:2025scf, Cacciapaglia:2006pk}.

In this paper, we attempt to provide an extension of the oblique parameters to models with a gauge sector extended by a dark Abelian symmetry $U(1)_D$, which we call the dark $U(1)$ models. 
In these models, a new gauge boson $Z_D$ couples to the SM fermions only via mixing with SM gauge bosons, which modifies the mass formulas and gauge couplings for the SM gauge bosons at tree level~\cite{Holdom:1985ag, Holdom:1990xp, Altarelli:1989qy, Burgess:1993vc, Babu:1997st}. 
When the mixing parameters are sufficiently small, this effect can be treated as a perturbative deviation from the SM and can be parametrized by the oblique parameters~\cite{Holdom:1990xp, Burgess:1993vc}. 
The tree-level oblique parameters have been used to investigate the EWPO constraint and the $W$ boson mass prediction~\cite{Thomas:2022gib, Cheng:2022aau, Harigaya:2023uhg, Bian:2017xzg, Lee:2022nqz, Lee:2021gnw, Davoudiasl:2023cnc, Jung:2023ckm, Bento:2023flt}.\footnote{The global fitting for the EWPO constraint in the dark $U(1)$ models are studied in Refs.\cite{Curtin:2014cca, Bertuzzo:2025ejw}.}

Although the formulations in Refs.~\cite{Holdom:1990xp, Burgess:1993vc} can also be applied to loop-level analyses, the one-loop contribution including the mixing effect has not been discussed.
However, such a contribution can be significant in some dark $U(1)$ models because the one-loop effect can be of lower order in the mixing than the tree-level one. 
In addition, the term linear in the mixing in the one-loop contribution makes the result different from that in the standard EW gauge theory. 
This novel contribution can drastically change the constraint on the models.

In this paper, we thoroughly investigate the oblique corrections in the dark $U(1)$ models in a model-independent way. 
To this end, we use the effective action method, which results in the renormalization scheme (RS) independent representation~\cite{Peskin:1990zt, Kennedy:1988sn}, to consider the four-fermion processes. 
We solve the Schwinger--Dyson (SD) equations for the transverse part of the gauge boson propagators and derive the scattering amplitude formulas including the oblique corrections at all orders of the perturbation. 
By expanding the formulas up to the one-loop level, we define the running parameters such as the masses, the gauge couplings, and the mixing parameters. 
After fixing the renormalized parameters by the on-shell conditions, we define the oblique parameters by approximating the two-point functions and show that they describes the oblique corrections when the mass scale of the loop diagrams is sufficiently high.\footnote{The effects mediated by the $Z_D$ boson cannot be included in the oblique parameters.  However, this hardly affect the observables at the $Z$ pole unless $Z_D$ has a nearly degenerate mass with the $Z$ boson~\cite{Harigaya:2023uhg}.}

We consider two classes of the dark $U(1)$ models: (a) dark photon models, where the gauge bosons are mixed by the non-diagonal kinetic terms, which is parametrized by $\varepsilon$~\cite{Holdom:1985ag}, and (b) dark $Z$ models, where another parameter $\varepsilon_Z^{}$ provides the additional source of the mixing~\cite{Davoudiasl:2012ag}.
The difference between two classes comes from how the gauge symmetry is spontaneously broken. 
In the latter class, we examine two RSs for $\varepsilon_Z$, which are interchangeable through an RS conversion. 
Our analysis is quite general and can be used in various dark $U(1)$ models. 
Our main results are presented in Eqs.~(\ref{eq: STU_DP}), (\ref{eq: STU_DZ_A}), and (\ref{eq: STU_DZ_B}), and their numerical impacts are shown in Figs.~\ref{fig: DP}, \ref{fig: DZ_A}, and \ref{fig: DZ_B}, respectively.

This paper is structured as follows. 
In Sec.~\ref{sec: models}, we explain the two classes of models considered in this paper.
In Sec.~\ref{sec: renormalized_amplitudes}, we solve the SD equation and derive the RS-independent amplitudes for the four-fermion processes with the full oblique corrections. 
We also define the running parameters at the one-loop level. 
In Sec.~\ref{sec: renormalization}, we consider the one-loop renormalization of the gauge couplings and the mixing parameters.  
Finally, the one-loop formulas for oblique parameters in each class of models are presented in Sec.~\ref{sec: STUparams}. 
As an example, we consider the effect of the dark isospin doublet and show the current EWPO constraint on the mixing parameters and the mass of $Z_D$. 
Sec.~\ref{sec: conclusions} summarizes our findings. 
Appendix~\ref{app: darkZ_model} shows an example of the dark $Z$ models. 
Appendix~\ref{app: absorptive_part} discusses the effect of the absorptive part of the two-point functions. 
In Appendix~\ref{app: divergence_cancellation}, we prove the finiteness of the running parameters in a general way.

%%%%%%%%%%%%%%%%%%%%%%%%%%%%%%%%%%%%%%%%%%%%%%%%%%
\section{Dark $U(1)$ models}
\label{sec: models}
%%%%%%%%%%%%%%%%%%%%%%%%%%%%%%%%%%%%%%%%%%%%%%%%%%

In this section, we focus on the gauge sector of the dark $U(1)$ models.  To facilitate the discussion, we will introduce two mixing parameters, $\varepsilon$ and $\varepsilon_Z^{}$, and discuss how they modify the gauge interactions of the SM fermions. 
All parameters and fields should be construed as bare quantities although no subscript such as $0$ or $B$ is used.

Let $\hat{Z}_D^\mu$ be the dark gauge field associated with $U(1)_D$ symmetry. 
The SM particles carry no dark charges and do not couple to $\hat{Z}_D^\mu$. 
The dark gauge field has a non-canonical kinetic term due to the mixing with the gauge field $\hat{B}^\mu$ for the hypercharge symmetry $U(1)_Y$~\cite{Holdom:1985ag}:
\begin{align}
\mathcal{L}_{\text{kin}}
= - \frac{ 1 }{ 4 } \hat{B}^{\mu\nu} \hat{B}_{\mu\nu}
+ \frac{ \varepsilon }{ 2  } \hat{B}_{\mu\nu} \hat{Z}^{\mu\nu}_{D} 
- \frac{ 1 }{ 4 } \hat{Z}_{D\mu\nu} \hat{Z}_D^{\mu\nu}, 
\end{align}
where $\hat{X}^{\mu\nu} = \partial^\mu \hat{X}^\nu - \partial^\nu \hat{X}^\mu$ for $X = B$ and $Z_D$, and $\varepsilon$ is the kinetic mixing parameter. 
The kinetic terms are diagonalized to the canonical form by the following $GR(2,R)$ transformation:
\begin{align}
\begin{pmatrix}
\hat{B}^\mu \\
\hat{Z}_D^\mu \\
\end{pmatrix}
=
\begin{pmatrix}
1 & \eta \, \varepsilon\\
0 & \eta \\
\end{pmatrix}
\begin{pmatrix}
\tilde{B}^\mu \\
\tilde{Z}_D^\mu \\
\end{pmatrix}, 
\end{align}
where $\eta = 1/\sqrt{1 - \varepsilon^2}$.

In such models, the covariant derivative is given by
\begin{align}
D^\mu = &\ \partial^\mu 
+ i g_L^{} I^a \hat{W}^{a\mu} 
+ i g_Y^{} Y \tilde{B}^\mu
+ i \eta \Bigl( \varepsilon g_Y Y + g_D Q_D \Bigr) \tilde{Z}_D^\mu, 
\end{align}
where $I^a$ ($a=$1,2,3) is the $a$-th component of the weak isospin, $Y$ is the hypercharge, $Q_D$ is the dark charge, $\hat{W}_\mu^a$ are the weak gauge bosons, and $g_L$, $g_Y$, and $g_D$ are the gauge couplings of $SU(2)_L$, $U(1)_Y$, and $U(1)_D$, respectively. 
Here, we have omitted the QCD term, which is identical to that in the SM.  This covariant derivative leads to the following current interactions among the SM fermions and the gauge bosons:
\begin{align}
\mathcal{L}_{\text{current}}
= & - \frac{ g_L }{ \sqrt{2} } \Bigl( W^{+}_\mu J_\mathrm{CC}^\mu + \mathrm{h.c.} \Bigr)
	- e A_\mu J_\mathrm{EM}^\mu
	\nonumber \\[5pt]
	& - g_Z \tilde{Z}_\mu \Bigl( J_3^\mu - \sin^2 \theta J_\mathrm{EM}^\mu \Bigr)
	-  \eta \varepsilon g_Z \sin \theta  \tilde{Z}_{D\mu} \Bigl(J_\mathrm{EM}^\mu -  J^\mu_3 \Bigr), 
\end{align}
where $g_Z = \sqrt{ g_L^2 + g_Y^2 }$, $\theta = \tan^{-1} (g_Y/g_L)$, $e = g_L \sin \theta$, $J_3^\mu$ is the fermion current induced by $I^3$, and $J_\mathrm{CC}^\mu$ and $J_\mathrm{EM}^\mu$ are the charged weak and electromagnetic (EM) currents defined as in the SM, respectively. The gauge bosons are defined by 
\begin{align}
W_\mu^\pm = \frac{ 1 }{ \sqrt{2} } \Bigl(\hat{W}_\mu^1 \mp i \hat{W}_\mu^2 \Bigr), \quad 
\begin{pmatrix}
\tilde{Z}_\mu \\
A_\mu
\end{pmatrix}
= 
\begin{pmatrix}
c_\theta & - s_\theta \\
s_\theta & c_\theta \\
\end{pmatrix}
\begin{pmatrix}
\hat{W}^3_\mu \\
\tilde{B}_\mu \\
\end{pmatrix}, 
\end{align}
where $s_\theta = \sin \theta$ and $c_\theta = \cos \theta$.\footnote{In the following, we use similar abbreviations for the trigonometric functions: $s_\chi = \sin \chi$, $c_\chi = \cos \chi$, and $t_\chi = \tan \chi$ for any angle $\chi$.}

The neutral gauge boson $A^\mu$ couples to $J_\mathrm{EM}^\mu$ and thus represents the photon. 
The other gauge bosons $\tilde{Z}^\mu$ and $\tilde{Z}_{D}^\mu$ are not mass eigenstates in general. 
The off-diagonal mass is induced by the kinetic mixing $\varepsilon$ and other sources related to the Higgs sector of the model as discussed below. 
In this paper, we consider two classes of models for their mass matrix.

%%%%%%%%%%%%%%%%%%%%%%%%%%%%%%%%%%%%%%%%%%%%%%%%%%
\subsection{Dark photon models}
\label{subsec: def_dark_photon_models}
%%%%%%%%%%%%%%%%%%%%%%%%%%%%%%%%%%%%%%%%%%%%%%%%%%

In this subsection, we consider dark photon models, where the gauge symmetry breaking is caused by the SM Higgs doublet $\phi$ and dark singlet scalars. 
For simplicity, we consider just one dark singlet $S$ with the dark charge $q_s$. 
The $\phi$ and $S$ fields acquire the following vacuum expectation values (VEVs);
\begin{align}
\left<\phi \right> 
= 
\frac{ 1 }{ \sqrt{2} } \begin{pmatrix}0 \\ v\end{pmatrix}, 
\quad 
\left< S \right> 
= 
\frac{ v_S }{ \sqrt{2} }.  
\end{align}
Then, the mass matrix for $\tilde{Z}^\mu$ and $\tilde{Z}_D^\mu$ are given by
\begin{align}
\label{eq: mass_matrix_DP}
M_V^2 = 
\begin{pmatrix}
\tilde{m}_Z^2 & - \tilde{m}_Z^2 \eta \varepsilon s_\theta \\
 - \tilde{m}_Z^2 \eta \varepsilon s_\theta & \tilde{m}_D^2 + \eta^2 \varepsilon^2 s_\theta^2 \\
\end{pmatrix}, 
\end{align}
where 
\begin{align}
\tilde{m}_Z^2 = \frac{ g_Z^2 }{ 4 } v^2, \quad 
\tilde{m}_D^2 = q_s^2 \eta^2 g_D^2 v_S^2.  
\end{align}
The mass eigenstates $Z$ and $Z_D$ bosons are then given by
\begin{align}
\begin{pmatrix}
Z^\mu \\
Z_D^\mu \\
\end{pmatrix}
= 
\begin{pmatrix}
\cos \xi & - \sin \xi \\
\sin \xi & \cos \xi \\
\end{pmatrix}
\begin{pmatrix}
\tilde{Z}^\mu \\
\tilde{Z}_D^\mu \\
\end{pmatrix}. 
\end{align}
The mixing angle $\xi$ satisfies 
\begin{align}
\label{eq: mixing_angle_DP}
\sin 2 \xi = \frac{2 \tilde{m}_Z^2 \eta \varepsilon s_\theta }{ m_Z^2 - m_D^2 }, 
\end{align}
where $m_Z$ and $m_D$ are the masses of the $Z$ and $Z_D$ bosons, respectively.
With a small kinetic mixing, the difference between $\tilde{m}_Z^2$ ($\tilde{m}_D^2$) and $m_Z^2$ ($m_D^2$) is of $\mathcal{O}(\varepsilon^2)$. 

In terms of $\xi$, the current interactions are given by
\begin{align}
\label{eq: current_interaction}
\mathcal{L}_{\text{current}}
= & - \frac{ g_L }{ \sqrt{2} } \Bigl( W^{+}_\mu J_\mathrm{CC}^\mu + \mathrm{h.c.} \Bigr)
	- e A_\mu J_\mathrm{EM}^\mu
	\nonumber \\[5pt]
	& - Z_\mu \Bigl\{ g_Z (c_\xi + s_\xi \eta \varepsilon s_\theta ) J_\mathrm{NC}^\mu 
		- e s_\xi \eta \varepsilon c_\theta J_\mathrm{EM}^\mu \Bigr\}
	\nonumber \\[5pt]
	& - Z_{D\mu} \Bigl\{ g_Z (s_\xi - c_\xi \eta \varepsilon s_\theta) J_\mathrm{NC}^\mu 
		+ e c_\xi \eta \varepsilon c_\theta J_\mathrm{EM}^\mu \Bigr\}, 
\end{align}
where $J_\mathrm{NC}^\mu = J_3^\mu - s_\theta^2 J_\mathrm{EM}^\mu$.\footnote{The kinetic mixing is often defined as $\varepsilon \hat{B}_{\mu\nu} \hat{Z}_D^{\mu\nu}/(2c_\theta)$, by which $Z_D$ has a simple EM coupling $- e \varepsilon Z_{D\mu} J_\mathrm{EM}^\mu$ for small $\varepsilon$. Here, we do not employ this convention because this simplification works only at tree level.}  
In the limit of no mixing ($\varepsilon \to 0$), the current interactions coincide with the SM ones.

Consequently, five independent parameters are required to describe the current interactions in the dark photon models: $g_L$, $g_Y$, $\varepsilon$, $v$, and $v_S$. 
In this paper, we use the following input parameters:
\begin{align}
\alpha, \quad G_F, \quad M_Z, \quad M_D, \quad \text{and} \quad \xi, 
\end{align}
where $\alpha$ is the fine structure constant, $G_F$ is the Fermi constant, and $M_Z$ and $M_D$ are the pole masses of the $Z$ and $Z_D$ bosons, respectively.
In this scheme, $\varepsilon$ is not a free parameter and is determined by the value of $\xi$ as discussed Sec.~\ref{sec: renormalization}.

We note that the explicit formula for $\tilde{m}_D^2$ is irrelevant in our discussions. 
The important point is that it is independent of the other input parameters.  
Therefore, this class of models can include multiple dark singlet scalars with nonzero VEVs in general. 
Also, it can include any other particle that do not contribute to the breaking of the gauge symmetries.

%%%%%%%%%%%%%%%%%%%%%%%%%%%%%%%%%%%%%%%%%%%%%%%%%%
\subsection{Dark $Z$ models}
\label{subsec: def_dark_Z_models}
%%%%%%%%%%%%%%%%%%%%%%%%%%%%%%%%%%%%%%%%%%%%%%%%%%

In models that include scalar fields with nonzero VEVs breaking both EW and dark symmetries, the gauge boson mass matrix $M_V^2$ has additional off-diagonal terms in general. 
We refer to such models as dark $Z$ models and parametrize the mass matrix as
\begin{align}
\label{eq: mass_matrix_DZ}
M_V^2 = 
\begin{pmatrix}
\tilde{m}^2_Z & - \tilde{m}^2_Z \eta (\varepsilon s_\theta + \varepsilon_Z) \\[10pt]
- \tilde{m}^2_Z \eta (\varepsilon s_\theta + \varepsilon_Z)  
	& \tilde{m}_D^2 + \eta^2 \varepsilon^2 s_\theta^2  \\
\end{pmatrix}. 
\end{align}
The additional parameter $\varepsilon_Z$ denotes the mass mixing in such models.  
The simplest example of the dark $Z$ models is the model with a new isospin doublet scalar with the dark charge, as shown in Appendix~\ref{app: darkZ_model}. 
This model was proposed in Ref.~\cite{Davoudiasl:2012ag} and has been studied in Refs.~\cite{Davoudiasl:2023cnc, Davoudiasl:2012qa, Lee:2013fda, Davoudiasl:2013aya, Davoudiasl:2014mqa, Davoudiasl:2014kua, Xu:2015wja, Davoudiasl:2015bua, San:2022uud, Goyal:2022vkg, Datta:2022zng, Jung:2023ckm, Sun:2023kfu, Cheng:2024hvq, Bertuzzo:2025ejw}. 
Extensions of the simplest model have also been discussed in Refs.~\cite{Chen:2025jyn, Bian:2017xzg, Lee:2022nqz, Lee:2021gnw, Bento:2023flt}. 
In the following discussion, the explicit formula of $\varepsilon_Z$ is irrelevant as long as it is independent of $\varepsilon$.

The mixing angle $\xi$ now satisfies 
\begin{align}
\label{eq: mixing_angle_DZ}
\sin 2 \xi = \frac{2 \tilde{m}_Z^2 \eta (\varepsilon s_\theta + \varepsilon_Z) }{ m_Z^2 - m_D^2 }. 
\end{align}
The current interactions have the same form as in the dark photon models and coincide with the SM ones in the no mixing limit ($\varepsilon \to 0$ and $\varepsilon_Z \to 0$). 

The dark $Z$ models have six free parameters to describe the current interactions. 
In this paper, we consider two ways of choosing the input parameters:
\begin{align}
\text{(i)} & \quad \alpha, \quad G_F, \quad M_Z, \quad M_D, \quad \xi, \quad \varepsilon_Z, \\[5pt]
\text{(ii)} & \quad \alpha, \quad G_F, \quad M_Z, \quad M_D, \quad \xi, \quad \varepsilon.
\end{align}
The first scheme is convenient for a direct comparison with the dark photon models because they coincide at the one-loop level as $\varepsilon_Z \to 0$ (the dark photon limit).
On the other hand, in the second scheme, $\varepsilon_Z$ is determined by the mixing angle $\xi$, and $\varepsilon$ is independent of $\xi$. 
This scheme would be more natural in the dark $Z$ models because the dependence on $\varepsilon_Z$ always appears only through $\xi$ via Eq.~(\ref{eq: mixing_angle_DZ}).

%%%%%%%%%%%%%%%%%%%%%%%%%%%%%%%%%%%%%%%%%%%%%%%%%%
\section{Oblique corrections in four-fermion processes and running parameters}
\label{sec: renormalized_amplitudes}
\setcounter{equation}{0}
%%%%%%%%%%%%%%%%%%%%%%%%%%%%%%%%%%%%%%%%%%%%%%%%%%

We now discuss the oblique corrections in the four-fermion processes. 
We consider only the light fermions as the external lines and neglect their masses, so that only the transverse modes of the gauge bosons contribute to the processes. 
In this section, all the bare quantities are represented with the subscript 0.

%%%%%%%%%%%%%%%%%%%%%%%%%%%%%%%%%%%%%%%%%%%%%%%%%%
\subsection{Amplitudes for four-fermion processes with oblique corrections}
%%%%%%%%%%%%%%%%%%%%%%%%%%%%%%%%%%%%%%%%%%%%%%%%%%

We follow the RS-independent way using the effective action according to Refs.~\cite{Kennedy:1988sn, Peskin:1990zt}. 
Denoting the transverse part of the propagator of gauge bosons $V$ and $V^\prime$ by $G_{VV^\prime}$, the scattering amplitudes of charged-current and neutral-current processes are then given by
\begin{align}
\label{eq: loop_MCC}
\mathcal{M}_\mathrm{CC} = & \frac{e_0^2 }{ 2 s_{\theta0}^2} I_+ I_- G_{WW}, \\[5pt]
\label{eq: loop_MNC}
\mathcal{M}_\mathrm{NC} = & \ e_0^2 QQ^\prime G_{AA} 
	+ e_0 \Bigl[Q v_Z^{f^\prime} + v_{Z}^f Q^\prime \Bigr] G_{ZA} 
    + e_0 \Bigl[Q v_D^{f^\prime} + v_{D}^f Q^\prime \Bigr] G_{DA} 
    \nonumber \\[5pt]
	& 	%
	+ v_Z^f v_Z^{f^\prime} G_{ZZ} 
    + v_D^f v_D^{f^\prime} G_{DD} 
    + \Bigl[ v_Z^f v_D^{f^\prime} + v_D^f v_Z^{f^\prime} \Bigr] G_{ZD}, 
\end{align}
where the fermion bilinears are omitted, $Q$ ($Q^\prime$) is the electric charges of an external fermion $f$ ($f^\prime$), $I_\pm$ are the isospin raising and lowering matrices, and $W$, $A$, $Z$, and $D$ represent the $W$ boson, the photon, the $Z$ boson, and the $Z_D$ boson, respectively. 
The vertex factors of the couplings between fermions $f$ and $Z$ ($Z_D$) boson are denoted by $v_Z^f$ ($v_D^f$), which are given by
\begin{align}
& v_Z^f = \frac{e_0}{s_{\theta 0}^{} c_{\theta 0}^{}} \Bigl\{ 
	c_{\xi0}^{} (I_3 - s_{\theta0}^2Q) - s_{\xi0}^{} \eta_0 \varepsilon_0 s_{\theta0}^{} (Q - I_3) \Bigr\}, \\[5pt]
& v_D^f = \frac{e_0}{s_{\theta0}^{} c_{\theta0}^{} } \Bigl\{ 
	s_{\xi0}^{} (I_3 - s_{\theta0}^2Q) + c_{\xi0}^{} \eta_0 \varepsilon_0 s_{\theta0}^{} (Q - I_3) \Bigr\}. 
\end{align}

At the leading order, the off-diagonal propagators $G_{ZA}$, $G_{DA}$, and $G_{ZD}$ vanish, and the diagonal ones are given by
\begin{align}
D_{VV} = \frac{ 1 }{ q^2 - m_{V0}^2 }, \quad V = W, A, Z, \text{ or } D, 
\end{align}
where $q^2$ is the squared external momentum, $m_{A0}^{} = 0$, $m_{W0}^2 = g_{L0}^2 v_0^2/4$, and $m_{Z0}^{2}$ and $m_{D0}^{2}$ are the eigenvalues of the mass matrix~(\ref{eq: mass_matrix_DP}) or (\ref{eq: mass_matrix_DZ}).

To include the oblique corrections, we consider the 1PI two-point functions of the gauge bosons, whose transverse parts are denoted by $\Pi_{VV^\prime}(q^2)$ and solve the SD equations for the exact propagators $G_{VV^\prime}$. 
For the $W$ boson propagator, the SD equation is given by
\begin{align}
G_{WW} = D_{WW} + D_{WW} \Pi_{WW} G_{WW}, 
\end{align}
which is the same as in the SM and is found to be 
\begin{align}
G_{WW} = \frac{ 1 }{ q^2 - m_{W0}^2 - \Pi_{WW} }. 
\end{align}
Thus, the charged-current amplitude is given by
\begin{align}
\label{eq: M_CC}
\mathcal{M}_\mathrm{CC} = \frac{ e_0^2 }{ 2 s_{\theta0}^2 } I_+ \frac{ 1 }{ q^2 - m_{W0}^2 - \Pi_{WW} } I_-. 
\end{align}

The equations for the neutral gauge bosons are more complicated than those in the SM:
\begin{align}
\left\{
\begin{array}{l}
 G_{AA} = D_{AA} + D_{AA} \Pi_{AA} G_{AA} + D_{AA} \Pi_{ZA} G_{ZA} 
	+ D_{AA} \Pi_{D A} G_{D A}, \\[5pt]
G_{ZZ} = D_{ZZ} + D_{ZZ} \Pi_{ZA} G_{ZA} + D_{ZZ} \Pi_{ZZ} G_{ZZ} 
	+ D_{ZZ} \Pi_{Z D} G_{Z D}, \\[5pt]
G_{ZA} = D_{ZZ} \Pi_{ZA} G_{AA} + D_{ZZ} \Pi_{ZZ} G_{ZA} + D_{ZZ} \Pi_{Z D} G_{D A}, \\[5pt]
G_{DD} = D_{DD} + D_{DD} \Pi_{D A} G_{D A}
	+ D_{DD} \Pi_{Z D} G_{Z D} 
	+ D_{DD} \Pi_{DD} G_{DD}, \\[5pt]
G_{D A} = D_{DD} \Pi_{D A} G_{AA} + D_{DD} \Pi_{ZD} G_{ZA} + D_{DD} \Pi_{DD} G_{D A}, \\[5pt]
G_{Z D} = D_{ZZ} \Pi_{ZA} G_{D A} + D_{ZZ} \Pi_{Z D} G_{DD} + D_{Z Z} \Pi_{Z Z} G_{Z D},
\end{array}
\right.
\end{align}
The solutions are then given by
\begin{align}
\label{eq: solution_of_SDeq}
\left\{
\begin{array}{ll}
\displaystyle{G_{AA} = \frac{ F_Z F_D - \Pi_{ZD}^2 }{ F_A F_Z F_D - \Sigma},} 
	& \displaystyle{G_{ZA} = \frac{ \Pi_{ZA}F_D +\Pi_{ZD}\Pi_{DA} }{  F_A F_Z F_D - \Sigma },}  \\[20pt]
\displaystyle{G_{ZZ} = \frac{ F_A F_D - \Pi_{DA}^2 }{ F_A F_Z F_D - \Sigma },} 
	& \displaystyle{G_{ZD} = \frac{ F_A \Pi_{ZD} + \Pi_{DA} \Pi_{ZA} }{ F_A F_Z F_D - \Sigma },}  \\[20pt]
\displaystyle{G_{DD} = \frac{ F_A F_Z - \Pi_{ZA}^2 }{ F_A F_Z F_D - \Sigma },}  
	& \displaystyle{G_{DA} = \frac{ \Pi_{DA} F_Z +\Pi_{ZD}\Pi_{ZA} }{  F_A F_Z F_D - \Sigma }, }
\end{array}
\right.
\end{align}
where 
\begin{align}
\begin{array}{l}
F_A = q^2 - \Pi_{AA}, \quad F_Z = q^2 - m_{Z0}^2 - \Pi_{ZZ}, \quad 
F_D = q^2 - m_{D0}^2 - \Pi_{DD}, \\[10pt]
\Sigma = F_A \Pi_{ZD}^2 + F_Z \Pi_{DA}^2 + F_D \Pi_{ZA}^2 + 2 \Pi_{ZA} \Pi_{DA} \Pi_{ZD}. 
\end{array}
\end{align} 
Using Eq.~(\ref{eq: solution_of_SDeq}), $G_{AA}$, $G_{ZA}$, and $G_{DA}$ can be expressed in terms of $G_{ZZ}$, $G_{ZD}$, and $G_{DD}$:
\begin{align}
\label{eq: solution_of_SDeq2}
\left\{
\begin{array}{l}
\displaystyle{
    G_{AA} = \frac{ 1 }{ F_A }
    \biggl[
      1 + \frac{ F_D\Pi_{ZA}^2 G_{ZZ} }{ F_AF_D - \Pi_{DA}^2  }
        + \frac{ F_Z \Pi_{DA}^2 G_{DD} }{ F_A F_Z - \Pi_{ZA}^2 }
        + \frac{ 2 \Pi_{ZA} \Pi_{ZD} \Pi_{DA} G_{ZD} }{ F_A \Pi_{ZD} + \Pi_{DA} \Pi_{ZA} }
    \biggr]
    }, \\[20pt]
\displaystyle{
    G_{ZA} = \frac{ G_{ZZ} }{ F_A } \frac{ F_A F_D \Pi_{ZA} }{ F_A F_D - \Pi_{DA}^2 } 
    + \frac{ \Pi_{ZD} \Pi_{DA} G_{ZD} }{ F_A \Pi_{ZD} + \Pi_{DA} \Pi_{ZA} }
    }, \\[20pt]
\displaystyle{
    G_{DA} = \frac{ G_{DD} }{ F_A } \frac{ F_A F_Z \Pi_{DA} }{ F_A F_Z - \Pi_{ZA}^2 } 
    + \frac{ \Pi_{ZD} \Pi_{ZA} G_{ZD} }{ F_A \Pi_{ZD} + \Pi_{DA} \Pi_{ZA} } 
    }.
\end{array}
\right.
\end{align}

By substituting these equations, the neutral-current amplitude is represented by using $F_A$, $G_{ZZ}$, $G_{DD}$, and $G_{ZD}$: 
\begin{align}
\label{eq: M_NC}
& \mathcal{M}_\mathrm{NC} = 
	 \frac{ e_0^2 Q Q^\prime }{ q^2 - \Pi_{AA} }
    + (V_Z^f, V_D^f)
	\begin{pmatrix}
	G_{ZZ} & G_{ZD} \\[7pt]
	G_{ZD} & G_{DD} \\
	\end{pmatrix}
	\begin{pmatrix}
	V_Z^{f^\prime} \\[7pt]
	V_D^{f^\prime}
	\end{pmatrix}, 
\end{align}
where 
\begin{align}
V_Z^f = v_Z^f + e_0 Q \frac{ \Pi_{ZA}  }{ q^2 - \Pi_{AA} }, \quad 
	V_D^f = v_D^f + e_0 Q \frac{ \Pi_{DA} }{ q^2 - \Pi_{AA} }. 
\end{align}
Other terms are canceled due to the following relations given by Eq.~(\ref{eq: solution_of_SDeq}):
\begin{align}
G_{ZZ} = \frac{ F_A F_D - \Pi_{DA}^2 }{ \Pi_{ZD} F_A + \Pi_{DA} \Pi_{ZA} }
	G_{ZD}, \quad
G_{DD} = \frac{ F_A F_Z - \Pi_{ZA}^2 }{ \Pi_{ZD} F_A + \Pi_{DA} \Pi_{ZA} }
	G_{ZD}. 
\end{align}

We note that no loop expansion has been used to derive Eqs.~(\ref{eq: M_CC}) and (\ref{eq: M_NC}). 
Hence, they are applicable to oblique corrections at all orders of the perturbation. 
Also, they are RS-invariant because they are expressed with only the bare parameters.

%%%%%%%%%%%%%%%%%%%%%%%%%%%%%%%%%%%%%%%%%%%%%%%%%%
\subsection{Running parameters}
\label{subsec: running_params}
%%%%%%%%%%%%%%%%%%%%%%%%%%%%%%%%%%%%%%%%%%%%%%%%%%

%Here, we define the running parameters. 
In the following, we consider $\Pi_{VV^\prime}$ at the one-loop level. 
For later convenience, we introduce $\tilde{\Pi}_{VV^\prime}(q^2)$ such that
\begin{align}
\Pi_{VV^\prime}(q^2) = \Pi_{VV^\prime}(0) + q^2 \tilde{\Pi}_{VV^\prime} (q^2).
\end{align}
We note that $\tilde{\Pi}_{VV^\prime}$ is identical to the derivative of $\Pi_{VV^\prime}$ only at $q^2 = 0$. 
Also, we neglect the absorptive parts of the two-point functions because they are irrelevant to the oblique corrections to the bare parameters. 
They are finite at the one-loop level and generate new forms of interactions as discussed in Appendix~\ref{app: absorptive_part}. 

Because of the EM gauge invariance, we expect that the two-point function with an external photon vanishes at $q^2 = 0$; 
\begin{align}
\label{eq: WT_identity}
\Pi_{AA}(0) = \Pi_{ZA}(0) = \Pi_{DA}(0) = 0.
\end{align}
However, na{\"i}ve loop calculations may violate these identities due to an artifact gauge dependence. 
As discussed in Ref.~\cite{Kennedy:1988sn}, this gauge dependence can be removed by including the gauge-dependent part of the vertex corrections, which is universal for external fermions, to define the modified two-point functions. 
This prescription can be systematically performed by using the pinch technique~\cite{Cornwall:1981zr, Papavassiliou:1989zd, Degrassi:1992ue, Duch:2018ucs}. 
In this paper, we do not discuss details of this issue because it is beyond the scope of this work. 
We just assume that $\Pi_{AV}(q^2)$ ($V=A$, $Z$, and $D$) are constructed in the gauge-invariant way so that Eq.~(\ref{eq: WT_identity}) is satisfied.

We begin with the photon contribution to the neutral-current processes; 
\begin{align}
\mathcal{M}_\mathrm{NC}^\gamma = \frac{ e_0^2 Q Q^\prime }{ q^2 }
    \left( \frac{ 1 }{ 1 - \tilde{\Pi}_{AA}(q^2) } \right), 
\end{align}
which has a pole at $q^2 = 0$. 
We note that Eq.~(\ref{eq: WT_identity}) makes the other terms in $\mathcal{M}_\mathrm{NC}$ regular at $q^2=0$. 
We thus define the running EM coupling $e_\ast(q^2)$ as 
\begin{align}
e_\ast^2(q^2) = \frac{ e_0^2 }{ 1 - \tilde{\Pi}_{AA}(q^2) } \simeq e_0^2 \Bigl( 1 + \tilde{\Pi}_{AA}(q^2) \Bigr),
\end{align}
the same as in the SM~\cite{Kennedy:1988sn, Peskin:1990zt}.

Next, we renormalize the propagators of the $Z$ and $Z_D$ bosons. 
The renormalized propagators $G^R_{VV^\prime}$ ($V, V^\prime = Z$ or $D$) are defined by 
\begin{align}
\begin{pmatrix}
G_{ZZ} & G_{ZD} \\[7pt]
G_{ZD} & G_{DD} \\
\end{pmatrix}
= 
\begin{pmatrix}
    Z_{ZZ}^{1/2} & Z_{ZD}^{1/2} \\
    Z_{DZ}^{1/2} & Z_{DD}^{1/2} \\
    \end{pmatrix}
	\begin{pmatrix}
	G_{ZZ}^R & G_{ZD}^R \\[7pt]
	G^R_{DZ} & G^R_{DD} \\
	\end{pmatrix}
    \begin{pmatrix}
    Z_{ZZ}^{1/2} & Z_{DZ}^{1/2} \\
    Z_{ZD}^{1/2} & Z_{DD}^{1/2} \\
    \end{pmatrix}, 
\end{align}
where we introduced the wave function renormalization (WFR) constants through
\begin{align}
\begin{pmatrix}
Z^0_\mu \\[7pt]
Z^0_{D\mu} 
\end{pmatrix}
= 
\begin{pmatrix}
Z_{ZZ}^{1/2} & Z_{ZD}^{1/2} \\[7pt]
Z_{DZ}^{1/2} & Z_{DD}^{1/2} \\
\end{pmatrix}
\begin{pmatrix}
Z_\mu \\[7pt]
Z_{D\mu} 
\end{pmatrix}. 
\end{align}
At the one-loop level, the renormalized propagators are given by
\begin{align}
\begin{array}{l}
\displaystyle{
	G_{ZZ}^R \simeq \frac{ Z_{ZZ}^{-1} }{ q^2 - m_{Z0}^2 - \Pi_{ZZ} }, \quad
	G_{DD}^R \simeq \frac{ Z_{DD}^{-1} }{ q^2 - m_{D0}^2 - \Pi_{DD} }
	}, \\[20pt]
\displaystyle{
	 G_{ZD}^R = G_{DZ}^R \simeq \frac{ \Pi_{ZD}^R(q^2) }{ (q^2 - m_{Z0}^2) (q^2 - m_{D0}^2) }
	 }, \\
\end{array}
\end{align}
where $\Pi_{ZD}^R$ is the renormalized $Z$-$Z_D$ 1PI function:
\begin{align}
\Pi_{ZD}^R(q^2) = \Pi_{ZD}(q^2) - Z_{DZ}^{1/2} (q^2 - m_{D0}^2) - Z_{ZD}^{1/2} (q^2 - m_{Z0}^2). 
\end{align}

The wave function and the mass are renormalized by using the on-shell conditions.
In this case, the pole masses $M_Z$ and $M_{D}$ are defined such that
\begin{align}
M_Z^2 = m_{Z0}^2 + \Pi_{ZZ}(M_Z^2), \quad 
M_{D}^2 = m_{D0}^2 + \Pi_{DD}(M_{D}^2), 
\end{align}
and $Z_{ZZ}$ and $Z_{DD}$ are fixed by
\begin{align}
Z_{VV} = \left( 1 - \frac{ \mathrm{d} \Pi_{VV}}{ \mathrm{d} q^2 } \biggr|_{q^2 = M_V^2} \right)^{-1} 
    \simeq 1 + \Pi^\prime_{VV}(M_V^2), 
\end{align}
for $V = Z$ and $D$, where $\Pi^\prime_{VV^\prime}(q^2)$ are the derivative of $\Pi_{VV^\prime}(q^2)$ with respect to $q^2$.
In addition, we impose the requirement that $G_{ZD}^R$ is regular at $q^2 = M_Z^2$ and $M_{D}^2$, leading to
\begin{align}
Z_{DZ}^{1/2} \simeq \frac{ \Pi_{ZD}(M_Z^2) }{ M_Z^2 - M_{D}^2 }, \quad 
Z_{ZD}^{1/2} \simeq - \frac{ \Pi_{ZD}(M_{D}^2) }{ M_Z^2 - M_{D}^2 }, 
\end{align}
where we have replaced $m_{Z0}$ and $m_{D0}$ with $M_Z$ and $M_D$, respectively, because the difference is of higher order. 
Using these counterterms, we can see that $\Pi_{ZD}^R(q^2)$ is finite.

Combining the above results, the renormalized propagators are given by 
\begin{align}
\begin{pmatrix}
G_{ZZ}^R & G_{ZD}^R \\[10pt]
G^R_{DZ} & G^R_{DD} \\
\end{pmatrix}
\simeq 
\begin{pmatrix} 
\displaystyle{ \frac{ 1 }{ q^2 - M_{Z\ast}^2(q^2)  } } & 
    \displaystyle{ \frac{ \Pi_{ZD}^R(q^2) }{ (q^2 - M_Z^2) (q^2 - M_D^2) } } \\[20pt]
\displaystyle{ \frac{ \Pi_{ZD}^R(q^2) }{ (q^2 - M_Z^2) (q^2 - M_D^2) } }
    & \displaystyle{ \frac{ 1 }{ q^2 - M_{D\ast}^2  } } 
\end{pmatrix},
\end{align}
where we have used the running masses defined by
\begin{align}
M_{V\ast}^2 (q^2) = M_V^2 + \Pi_{VV}(q^2) - \Pi_{VV}(M_V^2)
	- (q^2 - M_V^2) \Pi_{VV}^\prime (M_V^2), 
\end{align}
for $V = Z$ and $D$. 
The running masses satisfy $M_{V\ast}^2 (M_V^2) = M_V^2$. 

This propagator matrix is diagonalized by the momentum-dependent angle $\zeta (q^2)$; 
\begin{align}
\begin{pmatrix}
\cos \zeta & -\sin \zeta \\
\sin \zeta & \cos \zeta
\end{pmatrix}
\begin{pmatrix}
G_{ZZ}^R & G_{ZD}^R \\[7pt]
G^R_{DZ} & G^R_{DD} \\
\end{pmatrix}
\begin{pmatrix}
\cos \zeta & \sin \zeta \\
- \sin \zeta & \cos \zeta
\end{pmatrix}
\simeq 
\begin{pmatrix} 
\displaystyle{ \frac{ 1 }{ q^2 - M_{Z\ast}^2  } } & 
    0 \\[15pt]
0   & \displaystyle{ \frac{ 1 }{ q^2 - M_{D\ast}^2 } }
\end{pmatrix}, 
\end{align}
where
\begin{align}
\label{eq: zeta_def}
\sin 2 \zeta(q^2) = - \frac{2 \Pi_{ZD}^R(q^2)}{M_Z^2 - M_{D}^2}. 
\end{align}
Using the above quantities, we obtain the diagonal form of the neutral-current amplitude: 
\begin{align}
\label{eq: neutral amplitudes}
\mathcal{M}_\mathrm{NC} = 
	\frac{ e_\ast^2 Q Q^\prime }{ q^2 }
	+ V_{Z\ast}^f \frac{ 1 }{ q^2 - M_{Z\ast}^2 } V_{Z\ast}^{f^\prime} 
	+ V_{D\ast}^f \frac{ 1 }{ q^2 - M_{D\ast}^2  } V_{D\ast}^{f^\prime}.	
\end{align}
where $V_{Z\ast}^f$ and $V_{D\ast}^f$ are defined as
\begin{align}
\begin{array}{l}
V_{Z\ast}^f = \cos \zeta \Bigl( Z_{ZZ}^{1/2} V_Z^f + Z_{DZ}^{1/2} V_D^f \Bigr) - \sin \zeta \Bigl( Z_{ZD}^{1/2} V_Z^f + Z_{DD}^{1/2} V_D^f \Bigr), \\[15pt]
V_{D\ast}^f = \sin \zeta \Bigl( Z_{ZZ}^{1/2} V_Z^f + Z_{DZ}^{1/2} V_D^f \Bigr) + \cos \zeta \Bigl( Z_{ZD}^{1/2} V_Z^f + Z_{DD}^{1/2} V_D^f \Bigr).
\end{array} 
\end{align}

The remaining issue about $\mathcal{M}_\mathrm{NC}$ is to represent $V_{Z\ast}$ and $V_{D\ast}$ using the running parameters. 
For later convenience, we define the two-point functions for $\tilde{Z}_\mu$ and $\tilde{Z}_{D\mu}$ as 
\begin{align}
\tilde{\Pi}_{\tilde{Z}A} (q^2) = c_{\xi0} \tilde{\Pi}_{ZA} (q^2) 
	+ s_{\xi0} \tilde{\Pi}_{DA}(q^2), \quad 
	\tilde{\Pi}_{\tilde{D}A} (q^2) = -s_{\xi0} \tilde{\Pi}_{ZA} (q^2) 
	+ c_{\xi0} \tilde{\Pi}_{DA}(q^2). 
\end{align}
Since we expect that $V_{Z\ast}^f$ coincides with the SM one in the no mixing limit, the running weak mixing angle $s_{\theta\ast}^2$ is defined as 
\begin{align}
\label{eq: running_WMA}
s_{\theta \ast}^2(q^2) = s_{\theta0}^2 - s_{\theta0}^{} c_{\theta0}^{} \tilde{\Pi}_{\tilde{Z}A} (q^2). 
\end{align}
The other running parameters, the WFR constants $Z_{Z\ast}^{1/2}$ and $Z_{D\ast}^{1/2}$, the mixing angle $\xi_\ast$, and the kinetic mixing, are determined as follows by the requirements that they are divergence free and that $V_{Z\ast}^f$ and $V_{D\ast}^f$ can be represented with the same running parameters: 
\begin{align}
\label{eq: running_Zz}
& Z_{Z\ast}^{1/2} = Z_{ZZ}^{1/2} \biggl\{ 
    1 - \frac{ c_{\theta0}^2 - s_{\theta0}^2 }{ 2s_{\theta0} c_{\theta0} }\tilde{\Pi}_{ZA}
    - \frac{ 1 }{ 2 } \tilde{\Pi}_{AA}
    + \frac{ 1 }{ 2 } t_{\xi0}^{} \Bigl( X - t_{\theta0}^{} \tilde{\Pi}_{\tilde{D}A} \Bigr) 
    \biggr\}, \\[5pt]
& Z_{D\ast}^{1/2} = Z_{DD}^{1/2} \biggl( 
    1 - \frac{ 1 }{ 2 }\tilde{\Pi}_{AA}
    - \frac{ c_{\theta0}^2 - s_{\theta0}^2 }{ 2 s_{\theta0} c_{\theta0} } \tilde{\Pi}_{ZA} 
    + \frac{ 1 }{ 2 t_{\xi0}^{} } X 
    + \frac{ 1 }{ 2 t_{\xi0}^{} } t_{\theta0}^{} \tilde{\Pi}_{\tilde{D}A}
    \biggr), \\[5pt]
\label{eq: running_xi}
& \xi_\ast^{} = \xi_0^{} + \zeta - \frac{ Y }{ 2 } + \frac{ t_{\theta0} }{ 2 } \tilde{\Pi}_{\tilde{D}A}, \\[5pt]
\label{eq: running_epsilon}
& s_{\theta\ast} \eta_\ast \varepsilon_\ast = s_{\theta 0}^{}\eta_0^{} \varepsilon_0^{} 
    \biggl( 1 - \frac{ X }{ 2 s_{\xi0} c_{\xi0} } - \frac{ c_{\xi0}^2 - s_{\xi0}^2 }{ 2 s_{\xi0}^{} c_{\xi0}^{} } t_0 \tilde{\Pi}_{\tilde{D}A} 
    \biggr) + t_{\theta 0} \tilde{\Pi}_{\tilde{D}A}, 
\end{align}
where 
\begin{align}
\begin{array}{l}
X = Z_{DZ}^{1/2} + Z^{1/2}_{ZD}, \quad
Y = Z_{DZ}^{1/2} - Z^{1/2}_{ZD}. 
\end{array}
\end{align}
The expression of the running kinetic mixing $\varepsilon_\ast \eta_\ast$ is combined with $s_{\theta\ast}$ for the ease of later discussions. 
If necessary, $\varepsilon_\ast$ can be isolated using $s_{\theta\ast}^2$ in Eq.~(\ref{eq: running_WMA}) and $\eta_\ast = 1/\sqrt{1-\varepsilon_\ast^2}$.
By using these running quantities, $V_{Z\ast}^f$ and $V_{D\ast}^f$ are given by the same form as at tree level; 
\begin{align}
& V_{Z\ast}^f \simeq Z_{Z\ast}^{1/2} \frac{ e_\ast }{ s_{\theta\ast}^{} c_{\theta\ast}^{} }
	\biggl\{ 
	  c_{\xi\ast}^{}\bigl( I_3 - s_{\theta\ast}^2 Q \bigr) 
	- s_{\xi\ast}^{} s_{\theta\ast} (\eta\varepsilon)_\ast (Q-I_3)
	\biggr\}, \\[5pt]
& V_{D\ast}^f \simeq Z_{D\ast}^{1/2}
    \frac{ e_\ast }{ s_{\theta\ast}^{} c_{\theta\ast}^{} }
    \biggl\{ s_{\xi\ast}^{} \bigl(I_3 - s_{\theta\ast}^2 Q \bigr)
    + c_{\xi\ast}^{} s_{\theta\ast} \eta_\ast \varepsilon_\ast (Q-I_3) 
    \biggr\}. 
\end{align}

Finally, we consider the charged current amplitude in Eq.~(\ref{eq: M_CC}). 
The mass and WFR constant are renormalized by the on-shell condition, which leads to
\begin{align}
& M_W^2 = m_{W0}^2 + \Pi_{WW}(M_W^2), \\
& Z_W = \biggl( 1 - \frac{ \mathrm{d} \Pi_{WW} }{ \mathrm{d} q^2 } \biggr)^{-1} \simeq 1 + \Pi_{WW}^\prime (M_W^2). 
\end{align}
As in the SM, the running parameters are given by 
\begin{align}
\left\{
\begin{array}{l}
M_{W\ast}^2 = M_W^2 + \Pi_{WW}(q^2) - \Pi_{WW}(M_W^2) - (q^2 - M_W^2) \Pi^\prime_{WW} (M_W^2), \\[10pt]
\displaystyle{
    Z_{W\ast}^{1/2} = Z_{W}^{1/2} \biggl( 1 - \frac{ 1 }{ 2 } \tilde{\Pi}_{AA} - \frac{ c_{\theta0} }{ 2 s_{\theta 0} } \tilde{\Pi}_{\tilde{Z}A} \biggr)
    }.
\end{array}
\right.
\end{align}
By using them, the charged-current amplitude is given by 
\begin{align}
\mathcal{M}_\mathrm{CC} = \frac{ e_\ast^2 }{ 2 s_{\theta\ast}^2 }
    I_+ \frac{ Z_{W\ast} }{ q^2 - M_{W\ast}^2 } I_-. 
\end{align}

%%%%%%%%%%%%%%%%%%%%%%%%%%%%%%%%%%%%%%%%%%%%%%%%%%
\section{One-loop renormalization of couplings and mixing}
\label{sec: renormalization}
\setcounter{equation}{0}
%%%%%%%%%%%%%%%%%%%%%%%%%%%%%%%%%%%%%%%%%%%%%%%%%%

In this section, we consider the one-loop renormalization of the gauge couplings and mixing parameters, which is different between the dark photon models and the dark $Z$ models, as explained in Sec.~\ref{sec: models}.

%%%%%%%%%%%%%%%%%%%%%%%%%%%%%%%%%%%%%%%%%%%%%%%%%%
\subsection{Dark photon models}
%%%%%%%%%%%%%%%%%%%%%%%%%%%%%%%%%%%%%%%%%%%%%%%%%%

First, we discuss the dark photon models. 
As explained in Sec.~\ref{subsec: def_dark_photon_models}, there are five input parameters relevant to the four-fermion processes: $\alpha$, $G_F$, $M_Z$, $M_D$, and $\xi$. 
The masses $M_Z$ and $M_D$ have already been determined by the pole positions of the propagators in Sec.~\ref{subsec: running_params}.  The fine-structure constant $\alpha \simeq 1/137$ is determined by EM scattering at the Thomson limit;
\begin{align}
\alpha = \frac{ e_{\ast}(0)^2 }{ 4 \pi } \simeq 
    \frac{ e_0^2 }{ 4 \pi } \Bigl( 1 + \tilde{\Pi}_{AA}(0) \Bigr). 
\end{align}

The Fermi constant $G_F \simeq 1.166 \times 10^{-5}~\mathrm{GeV}^{-2}$ is determined by the charged-current amplitude at $q^2 = 0$:
\begin{align}
G_F  = \frac{e_\ast(0)^2 }{ 4 \sqrt{2} s_{\theta\ast}^{}(0)^2 } \frac{ 1 }{ M_{W\ast}^2(0) }
    \simeq G_F^0 \biggl( 1 - \frac{ \Pi_{WW}(0) }{ m_{W0}^2 } \biggr),  
\end{align}
where $G_F^0 = (\sqrt{2}v_0^2)^{-1}$.

To fix the input of the mixing angle $\xi$, we use the method in Ref.~\cite{Dittmaier:2023ovi, Denner:2018opp}, where we introduce a test fermion $\omega_d$ which has a dark charge $q_\omega$.
The limit of $q_\omega \to 0$ recovers the original theory. 
We impose the requirement that the ratio of the matrix elements of $Z \to \omega_d \bar{\omega}_d$ and $Z_D \to \omega_d \bar{\omega}_d$ equals to the tree-level formula $s_\xi/c_\xi$ in the limit of $q_\omega \to 0$. 
This results in the gauge-independent expression of $\xi$~\cite{Dittmaier:2023ovi}:
\begin{align}
\xi = & \ \xi_0 - s_{\xi0}^{} c_{\xi0}^{} \bigl( Z_{DD}^{1/2} - Z_{ZZ}^{1/2} \bigr) + s_{\xi0}^2 Z_{ZD}^{1/2} - c_{\xi0}^2 Z_{DZ}^{1/2}. 
\end{align}

All the input parameters have been fixed so far. 
Next, we discuss the dependent parameters: the weak mixing angle $\theta$ and the kinetic mixing $\varepsilon$. 
They are defined so as to satisfy the same relations at tree level; {\it i.e.},
\begin{align}
\sin 2 \theta = \sqrt{ \frac{ 4 \pi \alpha }{ \sqrt{2}G_F M^2 } }, \quad
\eta \varepsilon s_{\theta}^{} = \sin 2 \xi \biggl( \frac{ M_Z^2 - M_D^2 }{ 2 M^2 } \biggr), 
\end{align}
where $M^2 = c_\xi^2 M_Z^2 + s_\xi^2 M_D^2$.

Let $\delta A$ denote the one-loop contribution to a renormalized quantity $A$. 
By using 
\begin{align}
\delta s_\theta^2 = \frac{ 2 s_{\theta0}^2 c_{\theta0}^2 }{ c_{\theta0}^2 - s_{\theta0}^2 } \frac{ \delta \sin 2 \theta }{ \sin 2 \theta_0 }
= \frac{ s_{\theta0}^2 c_{\theta0}^2 }{ c_{\theta0}^2 - s_{\theta0}^2 }
    \biggl( \frac{ \delta \alpha }{ \alpha } - \frac{ \delta G_F }{ G_F } - \frac{ \delta M^2 }{ M^2 } \biggr), 
\end{align}
we can obtain 
\begin{align}
s_\theta^2 = s_{\theta0}^2 
    + \frac{ s_{\theta0}^2 c_{\theta0}^2 }{ c_{\theta0}^2 - s_{\theta0}^2 }
    \biggl( & \tilde{\Pi}_{AA}(0) + \frac{ \Pi_{WW}(0) }{ m_{W0}^2 }
        \nonumber \\[5pt]
        & - \frac{ c_{\xi0}^2 \Pi_{ZZ}(M_Z^2) + s_{\xi0}^2 \Pi_{DD}(M_D^2) }{ M^2 }
        + 2 s_{\xi0}^{} c_{\xi0}^{} \frac{ M_Z^2 - M_D^2 }{ M^2 } \delta \xi
    \biggr). 
\end{align}
Similarly, the renormalized kinetic mixing is given by
\begin{align}
\label{eq: EpsilonR_DP}
\eta \varepsilon s_\theta = 
    \eta_0^{} \varepsilon_0^{} s_{\theta0}^{}
    \biggl\{ & 1 + \frac{ c_{\xi0}^2 - s_{\xi0}^2 }{ s_{\xi0} c_{\xi0} } \delta \xi 
        + \frac{ \Pi_{ZZ}(M_Z^2) - \Pi_{DD}(M_D^2) }{ M_Z^2 - M_D^2 } 
        \nonumber \\[5pt]
        & - \frac{ 1 }{ M^2 }
        \biggl( c_{\xi0}^2 \Pi_{ZZ}(M_Z^2) 
            + s_{\xi0}^2 \Pi_{DD}(M_D^2) 
            - 2 s_{\xi0}^{} c_{\xi0}^{} (M_Z^2- M_D^2) \delta \xi 
        \biggr)\biggr\}.
\end{align}

Next, we discuss the $W$ boson mass. 
From the tree level relation, we can define the renormalized $W$ ``mass'' $m_W$ by
\begin{align}
m_W^2 = M^2 c_{\theta}^2 
= \frac{\pi \alpha}{\sqrt{2} G_F s_\theta^2}. 
\end{align}
However, $m_W$ is different from the pole mass $M_W$ due to the 1-loop correction. 
The difference $\Delta M_W^2 = M_W^2 - m_W^2$ is thus given by
\begin{align}
\frac{\Delta M_W^2}{m_W^2}
    = & \frac{ \Pi_{WW}(m_W^2) }{ m_W^2 }
    + \frac{ 1 }{ c_{\theta}^2 - s_\theta^2 }
    \biggl\{ s_\theta^2 \biggl(\tilde{\Pi}_{AA}(0) + \frac{ \Pi_{WW}(0)}{m_W^2} \biggr)
    \nonumber \\[5pt]
    & - \frac{ c_\theta^2 }{ M^2 }
            \Bigl(c_\xi^2 \Pi_{ZZ}(M_Z^2) + s_\xi^2 \Pi_{DD}(M_D^2)\Bigr)
    + 2 s_\xi c_\xi c_\theta^2 \biggl( \frac{ M_Z^2 - M_D^2 }{ M^2 } \biggr)\delta \xi
    \biggr\}, 
\end{align}
where $M_W$ and the bare quantities on the right-hand side can be replaced by $m_W$ and the corresponding renormalized quantities, respectively, because the difference is of higher order.

Using the above results, the one-loop running parameters are represented as follows for the dark photon models:
\begin{align}
& M_{V\ast}^2 = M_V^2 + \Bigl( \Pi_{VV}(q^2) - \Pi_{VV}(M_V^2) \Bigr) - (q^2 - M_V^2) \Pi_{VV}^\prime (M_V^2), \\[5pt]
& Z_{Z\ast} = 1 + \Pi_{ZZ}^\prime(M_Z^2) - \frac{ c_\theta^2 - s_\theta^2 }{ s_\theta c_\theta } \tilde{\Pi}_{\tilde{Z}A}(q^2) - \tilde{\Pi}_{AA}(q^2) 
    \nonumber \\[5pt]
    & \hspace{30pt} + t_\xi \biggl( \frac{ \Pi_{ZD}(M_Z^2)- \Pi_{ZD}(M_D^2) }{ M_Z^2 - M_D^2 } - t_{\theta}^{} \tilde{\Pi}_{\tilde{D}A}(q^2) \biggr), \\[5pt]
& Z_{D\ast} = 1 + \Pi_{DD}^\prime(M_D^2) - \frac{ c_\theta^2 - s_\theta^2 }{ s_\theta c_\theta } \tilde{\Pi}_{\tilde{Z}A}(q^2) - \tilde{\Pi}_{AA}(q^2) 
    \nonumber \\[5pt]
    & \hspace{30pt} + \frac{1}{t_\xi^{}} \biggl( \frac{ \Pi_{ZD}(M_Z^2)- \Pi_{ZD}(M_D^2) }{ M_Z^2 - M_D^2 } + t_{\theta}^{} \tilde{\Pi}_{\tilde{D}A}(q^2) \biggr), \\[5pt]
& e_\ast^2 = e^2\Bigl( 1 + \tilde{\Pi}_{AA}(q^2) - \tilde{\Pi}_{AA}(0) \Bigr), \\[5pt]
\label{eq: running_WMA_DP}
& s_{\theta\ast}^2 = s_\theta^2 
    - \frac{ s_\theta^2  c_\theta^2 }{ c_\theta^2 - s_\theta^2 } 
    \biggl\{ \tilde{\Pi}_{AA}(0) 
        + \frac{ \Pi_{WW}(0) }{ m_W^2 } 
        + \frac{ c_\theta^2 - s_\theta^2 }{ s_\theta^{} c_\theta^{} } \tilde{\Pi}_{\tilde{Z}A}(q^2)
        \nonumber \\[5pt]
        & \hspace{100pt} - \frac{ c_\xi^2 \Pi_{ZZ}(M_Z^2) + s_\xi^2 \Pi_{DD}(M_D^2) }{ M^2 } 
        + 2 s_\xi c_\xi \biggl( \frac{ M_Z^2 - M_D^2 }{ M^2 } \biggr) \delta \xi 
        \biggr\} , \\
& \xi_{\ast} = \xi + \frac{s_\xi c_\xi}{2} \Bigl( \Pi_{DD}^\prime (M_D^2) - \Pi_{ZZ}^\prime (M_Z^2) \Bigr)
    + \frac{ t_\theta }{2 } \tilde{\Pi}_{\tilde{D}A} (q^2)
    \nonumber \\[5pt]
    & \hspace{30pt} + \frac{ 1 }{ 2 (M_Z^2-M_D^2) } 
    \biggl\{ (c_\xi^2-s_\xi^2) \Bigl( \Pi_{ZD}(M_Z^2) - \Pi_{ZD}(M_D^2) \Bigr) - 2 \Pi_{ZD}^R(q^2) \biggr\}, \\[
    5pt]
\label{eq: running_epsilon_DP}
& (\eta \varepsilon s_\theta)_\ast = \eta \varepsilon s_\theta 
    \biggl\{ 1 - \frac{ c_\xi^2 - s_\xi^2 }{ s_\xi c_\xi } \delta \xi 
    - \frac{ \Pi_{ZZ}(M_Z^2) - \Pi_{DD}(M_D^2) }{ M_Z^2 - M_D^2 } 
    \nonumber \\[5pt]
    & \hspace{70pt} - \frac{ 1 }{ 2 s_\xi c_\xi } \biggl( \frac{ \Pi_{ZD}(M_Z^2) - \Pi_{ZD}(M_D^2) }{ M_Z^2 - M_D^2 } \biggr)
    - \frac{ c_\xi^2 - s_\xi^2 }{ 2 s_\xi c_\xi } t_\theta \tilde{\Pi}_{\tilde{D}A}(q^2)
    \nonumber \\[5pt]
    &\hspace{70pt}  + \frac{ c_\xi^2 \Pi_{ZZ}(M_Z^2) + s_\xi^2 \Pi_{DD}(M_D^2) - 2 s_\xi c_\xi (M_Z^2-M_D^2) \delta \xi }{ M^2 }
    \biggr\} 
    + t_\theta \tilde{\Pi}_{\tilde{D}A} (q^2), 
\end{align}
where $V= Z$, $D$, and $W$. 
In Appendix~\ref{app: divergence_cancellation}, we prove that the divergences are canceled in all these running parameters.

%%%%%%%%%%%%%%%%%%%%%%%%%%%%%%%%%%%%%%%%%%%%%%%%%%
\subsection{Dark $Z$ models} 
%%%%%%%%%%%%%%%%%%%%%%%%%%%%%%%%%%%%%%%%%%%%%%%%%%

Here, we discuss the one-loop renormalization in the dark $Z$ models. 
The difference from the dark photon models is in the definition of the mixing parameter $\varepsilon_Z^{}$. 
As commented in Sec.~\ref{subsec: def_dark_Z_models}, we consider two RSs for $\varepsilon_Z$.

%%%%%%%%%%%%%%%%%%%%%%%%%%%%%%%%%%%%%%%%%%%%%%%%%%
\subsubsection{Renormalization Scheme A: $\overline{MS}$ scheme for the mass mixing}
\label{sec: darkZ_schemeA}
%%%%%%%%%%%%%%%%%%%%%%%%%%%%%%%%%%%%%%%%%%%%%%%%%%

In this RS (RS-A), the tree-level relation (\ref{eq: mixing_angle_DZ}) is used to define $\varepsilon$, and the $\varepsilon_Z^{}$ parameter is determined by the $\overline{\mathrm{MS}}$ scheme. 
This scheme is convenient for a comparison with the dark photon models because they match in the limit of $\varepsilon_Z \to 0$ at one-loop level.

The running kinetic mixing in this scheme is then given by
\begin{align}
\label{eq: running_epsilon_DZ_A}
(\eta \varepsilon s_\theta)_\ast 
= (\eta \varepsilon s_\theta)_\ast^\mathrm{DP} 
+ \delta (\eta \varepsilon_Z)^{\overline{\mathrm{MS}}}, 
\end{align}
where $(\eta \varepsilon s_\theta)_\ast^\mathrm{DP}$ is the running kinetic mixing in the dark photon models in Eq.~(\ref{eq: running_epsilon_DP}). 
See Appendix~\ref{app: divergence_cancellation} for the explicit formula of the counterterm $\delta (\eta \varepsilon_Z)^{\overline{\mathrm{MS}}}$.
All the other running parameters are the same as in the dark photon models.

%%%%%%%%%%%%%%%%%%%%%%%%%%%%%%%%%%%%%%%%%%%%%%%%%%
\subsubsection{Renormalization Scheme B: Using $\xi$ to determine the mass mixing}
\label{sec: darkZ_schemeB}
%%%%%%%%%%%%%%%%%%%%%%%%%%%%%%%%%%%%%%%%%%%%%%%%%%

In this RS (RS-B), the mass mixing is renormalized by using Eq.~(\ref{eq: running_epsilon_DP}). To avoid confusion, we express the renormalized kinetic and mass mixing in this scheme by $\hat{\varepsilon}$ and $\hat{\varepsilon}_Z$, respectively. 

We define $\hat{\varepsilon}$ by using the running kinetic mixing in Eq.~(\ref{eq: running_epsilon}) at $q^2 = \mu^2$;  
\begin{align}
\label{eq: EpsilonR_DZB}
\hat{\eta}\hat{\varepsilon} = 
    \eta_0 \varepsilon_0 \biggl\{
        & 1 + \frac{ c_{\theta0}^{} }{ 2 s_{\theta0}^{} } \tilde{\Pi}_{\tilde{Z}A}(\mu^2) 
        - \frac{ 1 }{ 2 s_{\xi0} c_{\xi0} } \biggl( \frac{ \Pi_{ZD}(M_Z^2) - \Pi_{ZD}(M_D^2) }{ M_Z^2 - M_D^2 } \biggr) 
        \nonumber \\[5pt]
        & + \biggl( \frac{ 1 }{\eta_0 \varepsilon_0s_{\theta 0}^{}} - \frac{ c_{\xi0}^2 - s_{\xi0}^2 }{ 2 s_{\xi0}^{} c_{\xi0}^{} } \biggr) t_{\theta 0}^{} \tilde{\Pi}_{\tilde{D}A}(\mu^2)
    \biggr\}, 
\end{align}
where $\hat{\eta} = 1/\sqrt{1-\hat{\varepsilon}^2}$. 
We note that the right-hand side includes the running effect of $s_{\theta\ast}^2(\mu^2)$ in Eq.~(\ref{eq: running_epsilon}). 
The renormalized mass mixing is determined by the tree-level relation (\ref{eq: mixing_angle_DZ}):
\begin{align}
\hat{\varepsilon}_Z = - \hat{\varepsilon} s_\theta + \sin 2\xi \biggl( \frac{ M_Z^2 - M_D^2 }{ 2 M^2 } \biggr). 
\end{align}

In this scheme, the running kinetic mixing is given by
\begin{align}
\label{eq: running_epsilon_DZ_schemeB}
(\eta \varepsilon)_{\ast} = \hat{\eta}\hat{\varepsilon}
    \biggl( 1 + \frac{ 1 }{ 2 t_\theta } \Delta \tilde{\Pi}_{\tilde{Z}A}(q^2) + \frac{ c_\xi^2 - s_\xi^2 }{ 2 s_\xi c_\xi } t_\theta \Delta \tilde{\Pi}_{\tilde{D}A}(q^2) \biggr)
    + \frac{ 1 }{ c_\theta } \Delta \tilde{\Pi}_{\tilde{D}A}(q^2), 
\end{align}
where $\Delta \tilde{\Pi}_{\tilde{V}A}(q^2) = \tilde{\Pi}_{\tilde{V}A}(q^2) - \tilde{\Pi}_{\tilde{V}A}(\mu^2)$ for $V = Z$ and $D$. 
All the other running parameters are the same as in the dark photon models. 

Before closing this section, we comment on another possible way to renormalize the mass mixing. 
Since the mass mixing originates from the Higgs sector as discussed above, 
it can be fixed by renormalizing the scalar couplings. 
Clearly, this method strongly depends on the Higgs potential of the model. 
We thus do not discuss this possibility further to avoid the loss of generality.

%%%%%%%%%%%%%%%%%%%%%%%%%%%%%%%%%%%%%%%%%%%%%%%%%%
\section{The $S$, $T$, and $U$ parameters}
\label{sec: STUparams}
%%%%%%%%%%%%%%%%%%%%%%%%%%%%%%%%%%%%%%%%%%%%%%%%%%

In this section, we investigate how the oblique corrections change the EW observables and define the $S$, $T$, and $U$ parameters at the one-loop level. 
We neglect the contribution of the $Z_D$ mediation, the third term of Eq.~(\ref{eq: neutral amplitudes}), because its effect is suppressed in the observables at the $Z$ pole unless $M_D \simeq M_Z$.

%%%%%%%%%%%%%%%%%%%%%%%%%%%%%%%%%%%%%%%%%%%%%%%%%%
\subsection{Tree level formulas}
%%%%%%%%%%%%%%%%%%%%%%%%%%%%%%%%%%%%%%%%%%%%%%%%%%

First, we review the tree-level discussion. 
We define the effective weak mixing angle $\theta_\mathrm{eff}$ and the effective WFR constants $Z_{Z,\mathrm{eff}}$ and $Z_{W,\mathrm{eff}}$ to represent the tree-level amplitudes in the SM form; 
\begin{align}
& \mathcal{M}_\mathrm{NC}^\mathrm{tree} 
    = \frac{e^2}{q^2}QQ^\prime 
    + \frac{e^2}{ s_\mathrm{eff}^2 c_\mathrm{eff}^2 } (I_3 - s_\mathrm{eff}^2 Q) 
    \frac{ Z_{Z,\mathrm{eff}} }{ q^2 - M_Z^2 }  (I_3^\prime - s_\mathrm{eff}^2 Q^\prime), \\[5pt]
& \mathcal{M}_\mathrm{CC}^\mathrm{tree} = \frac{ e^2 }{ 2 s_{\mathrm{eff}}^2 }
    I_+ \frac{ Z_{W,\mathrm{eff}} }{ q^2 - m_W^2 } I_-. 
\end{align}
By comparing these equations and the tree-level formulas, we obtain
\begin{align}
\label{eq: def_effective_WMA}
& \sin^2 \theta_\mathrm{eff} = 
    \frac{ s_\theta^2 c_\xi + s_\xi \eta \varepsilon s_\theta }{ c_\xi + s_\xi \eta \varepsilon s_\theta }, \quad 
\cos^2 \theta_\mathrm{eff} = 1 - \sin^2 \theta_\mathrm{eff}, \\[5pt]
& Z_{Z,\mathrm{eff}} 
    = \frac{ s_\mathrm{eff}^2 c_\mathrm{eff}^2 }{ s_\theta^2 c_\theta^2 } Z_Z (c_\xi + s_\xi \eta \varepsilon s_\theta)^2
    = Z_Z c_\xi (c_\xi + s_\xi \eta \varepsilon s_\theta^{-1} ), \\[5pt]
\label{eq: def_effective_ZW}
& Z_{W,\mathrm{eff}} 
    = \frac{ s_\mathrm{eff}^2 }{ s_\theta^2} Z_W
    = Z_W \biggl( \frac{ c_\xi + s_\xi \eta \varepsilon s_\theta^{-1} }{ c_\xi + s_\xi \eta \varepsilon s_\theta } \biggr), 
\end{align}
where $s_\mathrm{eff} = \sin \theta_\mathrm{eff}$ and $c_\mathrm{eff} = \cos \theta_\mathrm{eff}$.
The tree-level WFR constants, $Z_Z = Z_W = 1$, is explicitly shown for later convenience.
We note that all the deviation from the SM are absorbed into $s_\mathrm{eff}^2$, $Z_{Z,\mathrm{eff}}$, $Z_{W,\mathrm{eff}}$, and $m_W^2 = c_\theta^2 M^2$.

We define the oblique parameters as in the standard EW gauge theory~\cite{Peskin:1990zt};
\begin{align}
\label{eq: STU_tree_def}
\left\{
\begin{array}{l}
\alpha S_\mathrm{tree} = 4 s_W^2 c_W^2 \Bigl(Z_{Z,\mathrm{eff}}-1\Bigr), \\[5pt]
\alpha T_\mathrm{tree} = \rho -1, \\[5pt]
\alpha U_\mathrm{tree} = 4 s_W^2 \Bigl( Z_{W,\mathrm{eff}} -1 \Bigr) - \alpha S_\mathrm{tree},
\end{array}
\right.
\end{align}
where $\rho$ is the ratio of the strengths of the charged weak current and the neutral weak current at zero momentum~\cite{Peskin:1990zt}. 
By expanding the effective quantities by the small mixing, 
the leading terms are given by the quadratics of the mixing parameters:
\begin{align}
\label{eq: STU_tree}
\left\{
\begin{array}{l}
\alpha S_\mathrm{tree} \simeq - 4 s_W c_W^2 \xi \bigl( \xi s_W - \varepsilon), \\[5pt]
\alpha T_\mathrm{tree} \simeq \xi^2 (r^2-2) + 2 \xi \varepsilon s_W, \\[5pt]
\alpha U_\mathrm{tree} \simeq 4 s_W^2 c_W^2 \xi^2, 
\end{array}
\right.
\end{align}
where $s_W = \sin \theta_W$, $c_W = \cos \theta_W$, and $\theta_W$ is the weak mixing angle in the SM defined as 
\begin{align}
\sin 2 \theta_W = \sqrt{ \frac{ 4 \pi \alpha }{ \sqrt{2 } G_F M_Z^2 } }. 
\end{align}
Since the mixing effect in $s_\mathrm{eff}^2$ and $m_W^2$ can also be described by the $S$, $T$, and $U$ parameters in the same form as in the standard EW gauge theory~\cite{Peskin:1990zt},   
all the mixing effect in the tree-level amplitudes can be described by three oblique parameters~\cite{Holdom:1990xp, Burgess:1993vc}.

%%%%%%%%%%%%%%%%%%%%%%%%%%%%%%%%%%%%%%%%%%%%%%%%%%
\subsection{One-loop formulas}
%%%%%%%%%%%%%%%%%%%%%%%%%%%%%%%%%%%%%%%%%%%%%%%%%%

Next, we extend the discussion to the one-loop level. 
We take into account the one-loop corrections of the order of $\hbar$, $\hbar \varepsilon$, and $\hbar \varepsilon_Z^{}$ so that we can neglect the interference between the tree-level deviations and the one-loop deviations. 
This enables us to linearize the deviation from the SM formula in any processes. 

By using the running parameters, the one-loop level amplitudes are given by
\begin{align}
\begin{array}{l}
\displaystyle{
    \mathcal{M}_\mathrm{NC}
    = \frac{e^2_\ast}{q^2}QQ^\prime 
    + \frac{e^2_\ast}{s_{\mathrm{eff}\ast}^2 c_{\mathrm{eff}\ast}^2 } (I_3 - s_{\mathrm{eff}\ast}^2 Q) \frac{ Z_{Z,\mathrm{eff}\ast} }{ q^2 - M_{Z\ast}^2 } (I_3^\prime - s_{\mathrm{eff}\ast}^2 Q^\prime) 
    }, \\[20pt]
\displaystyle{
    \mathcal{M}_\mathrm{CC} = \frac{ e_\ast^2 }{ 2 s_{\mathrm{eff}\ast}^2 }
    I_+ \frac{ Z_{W,\mathrm{eff}\ast} }{ q^2 - M_{W\ast}^2 } I_-
    }, 
\end{array}
\end{align}
where $s_\mathrm{eff\ast}^2$, $Z_{Z,\mathrm{eff}\ast}$, and $Z_{W,\mathrm{eff}\ast}$ are given by replacing the bare quantities in Eqs.~(\ref{eq: def_effective_WMA})-(\ref{eq: def_effective_ZW}) with the corresponding running parameters. 
Thus, the one-loop effects in the four-fermion processes are described by six functions of $q^2$: $e_\ast^2$, $M_{Z\ast}^2$, $M_{W\ast}^2$, $s_\mathrm{eff\ast}^2$, $Z_{Z,\mathrm{eff}\ast}$, and $Z_{W,\mathrm{eff}\ast}$. 

The oblique parameters are defined in the same way as at tree level but using the running parameters. 
Up to the linear order of the mixing, their one-loop terms are given by
\begin{align}
\label{eq: STUparams_general_rep}
\begin{array}{l}
\displaystyle{
    \Delta (\alpha S) \simeq 4 s_W^2 c_W^2 
    \biggl\{ \Delta Z_{Z\ast} 
        + \biggl( \frac{ \varepsilon }{ s_W } - 2 \xi \biggr) \Delta \xi_\ast
        + \frac{ \xi }{ s_W^2 } \Delta (\eta \varepsilon s_\theta)_\ast 
    \biggr\}
    }, \\[15pt]
\displaystyle{
    \Delta (\alpha T) \simeq 
    \Delta Z_{Z\ast} - \Delta Z_{W\ast} 
    + \frac{ \Delta M_{W\ast}^2 }{ m_W^2 }
    + \frac{ \Delta s_{\theta\ast}^2 }{ c_W^2 }
    - 2(\xi-\varepsilon s_W) \Delta \xi_\ast
    + 2\xi \Delta (\eta\varepsilon s_\theta)_\ast
    \biggr|_{q^2 = 0}
    }, \\[15pt]
\displaystyle{
    \Delta (\alpha U) \simeq 4 s_W^2 
    \Bigl( \Delta Z_{W\ast} - c_W^2 \Delta Z_{Z\ast}
        + 2\xi c_W \Delta \xi_\ast
    \Bigr)
    }. 
\end{array}
\end{align}
where $\Delta A$ represents the one-loop contribution in the parameter $A$. 
These formulas are applicable to both the dark photon models and dark $Z$ models. 

In the following, we will shift the oblique parameters so that they are zero in the SM limit to focus on their deviations from the SM. 
After such a shift, they still describe all the new physics effects because the deviations can be linearized at the order of perturbation we consider. 
Furthermore, for simplicity, we approximate the momentum dependence of the two-point functions as~\cite{Peskin:1990zt}
\begin{align}
\label{eq: expanded_Pi}
\Pi_{VV^\prime}(q^2) \simeq \Pi_{VV^\prime}(0) + q^2 \Pi_{VV^\prime}^\prime(0). 
\end{align}
This is valid as long as new particles in the loop diagrams are sufficiently heavy.
With this approximation, $\tilde{\Pi}_{VV^\prime}$ is equal to $\Pi_{VV^\prime}$, and we use $\Pi_{VV^\prime}^\prime$ consistently below. 

Beyond this point, the calculations are different between the dark photon models and the dark $Z$ models and also sensitive to the choice of RSs. We examine each case individually. 

%%%%%%%%%%%%%%%%%%%%%%%%%%%%%%%%%%%%%%%%%%%%%%%%%%
\subsubsection{Dark photon models}
%%%%%%%%%%%%%%%%%%%%%%%%%%%%%%%%%%%%%%%%%%%%%%%%%%

We first consider the dark photon models. 
By using Eq.~(\ref{eq: expanded_Pi}) with the expansions by the small mixing, the one-loop contributions in the oblique parameters are given by
\begin{align}
\label{eq: STU_DP}
\begin{array}{l}
\displaystyle{\alpha \Delta S_\mathrm{DP} \simeq 
    4 s_W^2 c_W^2 \biggl( 
        \Pi_{ZZ}^\prime - \frac{ c_W^2 - s_W^2 }{ s_W c_W } \Pi_{ZA}^\prime - \Pi_{AA}^\prime 
    + \frac{\varepsilon}{s_W}\frac{\Pi_{ZD}+M_Z^2\Pi_{ZD}^\prime}{M_Z^2 - M_D^2}
    \biggr),} \\[15pt]
\displaystyle{\alpha \Delta T_\mathrm{DP} \simeq 
    \frac{\Pi_{WW}}{m_W^2} - \frac{\Pi_{ZZ}}{M_Z^2}, } \\[15pt]
\displaystyle{\alpha \Delta U_\mathrm{DP} \simeq 
    4 s_W^2 \Bigl( 
        \Pi_{WW}^\prime  - c_W^2 \Pi_{ZZ}^\prime 
        - 2 s_W c_W \Pi_{ZA}^\prime - s_W^2 \Pi_{AA}^\prime 
        \Bigr),} 
\end{array}
\end{align}
where we have used $\xi \simeq \varepsilon s_W/(1-r^2)$, and all the two-point functions are evaluated at $q^2=0$. 
Here, the order of the mixing of the two-point functions are not fixed, and they are chosen so as to make the final results valid up to the linear order. 
Therefore, even though $\Delta T_\mathrm{DP}$ and $\Delta U_\mathrm{DP}$ have the same forms as in the standard EW gauge theory, their explicit formulas can be different due to the mixing. 

As explained above, all the one-loop effects in the four-fermion processes are included in the six running parameters: $e_\ast^2$, $M_{Z\ast}^2$, $M_{W\ast}^2$, $s_\mathrm{eff\ast}^2$, $Z_{Z,\mathrm{eff}\ast}$, and $Z_{W,\mathrm{eff}\ast}$. 
The approximation in Eq.~(\ref{eq: expanded_Pi}) leads to $e_\ast \simeq e$ and $M_{Z\ast} \simeq M_Z$. 
The WFR constants $Z_{Z,\mathrm{eff}\ast}$ and $Z_{W,\mathrm{eff}\ast}$ can be represented by the $S$ and $U$ parameters by their definitions. 
In addition, we can show that $s_{\mathrm{eff}\ast}^2$ and $M_{W\ast}^2$ can also be represented by using the oblique parameters as follows: 
\begin{align}
\label{eq: seff_DP_STU}
& s_{\mathrm{eff}\ast}^2 -s_W^2 = \frac{1}{c_W^2-s_W^2}
    \biggl( \frac{\alpha S_\mathrm{DP} }{4} - s_W^2 c_W^2 \alpha T_\mathrm{DP} \biggr), \\[5pt]
\label{eq: MW_DP_STU}
& M_{W\ast}^2 - M_Z^2 c_W^2 \simeq M_W^2 - M_Z^2 c_W^2 
    = M_Z^2 c_W^2 \biggl( -\frac{\alpha S_\mathrm{DP} }{ 2 (c_W^2 - s_W^2) }
        + \frac{ \alpha T_\mathrm{DP} }{ 1 - t_W^2 } 
        + \frac{ \alpha U_\mathrm{DP} }{ 4 s_W^2 } \biggr), 
\end{align}
where tree level effect is included. 
These are the same relations as in the standard EW gauge theory~\cite{Peskin:1990zt}. 
Consequently, all the new physics effect can be described by the oblique parameters if the new physics scale is high enough to satisfy Eq.~(\ref{eq: expanded_Pi}).

%%%%%%%%%%%%%%%%%%%%%%%%%%%%%%%%%%%%%%%%%%%%%%%%%%
\subsubsection{Dark $Z$ models using RS-A}
%%%%%%%%%%%%%%%%%%%%%%%%%%%%%%%%%%%%%%%%%%%%%%%%%%

Next, we consider the dark $Z$ models using RS-A discussed in Sec.~\ref{sec: darkZ_schemeA}.
The one-loop contributions to the oblique parameters are given by 
\begin{align}
& \alpha \Delta S_{\text{DZ-A}} \simeq 
    4 s_W^2 c_W^2 \biggl( 
        \Pi_{ZZ}^\prime - \frac{ c_W^2 - s_W^2 }{ s_W c_W } \Pi_{ZA}^\prime - \Pi_{AA}^\prime 
    + \frac{\varepsilon}{s_W}\frac{\Pi_{ZD}+M_Z^2\Pi_{ZD}^\prime}{M_Z^2 - M_D^2}
    \biggr)
    + 4 \xi c_W^2 \delta (\eta \varepsilon_Z)^{\overline{\mathrm{MS}}}, \nonumber \\[20pt]
\label{eq: STU_DZ_A}
& \alpha \Delta T_{\text{DZ-A}} \simeq 
    \frac{\Pi_{WW}}{m_W^2} - \frac{\Pi_{ZZ}}{M_Z^2}
    -2 \varepsilon_Z^{} 
    \biggl( \frac{ \Pi_{ZD} + M_Z^2 \Pi_{ZD}^\prime}{ M_Z^2 - M_D^2 }
    \biggr)
    + 2 \xi \delta (\eta \varepsilon_Z)^{\overline{\mathrm{MS}}}, \\[20pt]
& \alpha \Delta U_{\text{DZ-A}} \simeq 
    4 s_W^2 \Bigl( 
        \Pi_{WW}^\prime  - c_W^2 \Pi_{ZZ}^\prime 
        - 2 s_W c_W \Pi_{ZA}^\prime - s_W^2 \Pi_{AA}^\prime 
        \Bigr). \nonumber  
\end{align} 
It is straightforward to check that Eqs.~(\ref{eq: seff_DP_STU}) and (\ref{eq: MW_DP_STU}) also hold in this RS. 
Therefore, new physics effects from the oblique corrections can be described by only the oblique parameters as in the dark photon models.

%%%%%%%%%%%%%%%%%%%%%%%%%%%%%%%%%%%%%%%%%%%%%%%%%%
\subsubsection{Dark $Z$ models using RS-B}
%%%%%%%%%%%%%%%%%%%%%%%%%%%%%%%%%%%%%%%%%%%%%%%%%%

Finally, we consider the dark $Z$ models using RS-B discussed in Sec.~\ref{sec: darkZ_schemeB}.
The one-loop contributions to the oblique parameters are given by
\begin{align}
& \alpha \Delta S_{\text{DZ-B}} \simeq 
    4 s_W^2 c_W^2 \biggl( 
        \Pi_{ZZ}^\prime - \frac{ c_W^2 - s_W^2 }{ s_W c_W } \Pi_{ZA}^\prime - \Pi_{AA}^\prime 
    + \frac{\hat{\varepsilon}}{2s_W} \Pi_{ZD}^\prime 
    + \frac{1}{2s_Wc_W}\bigl(\hat{\varepsilon} s_W - 2\xi \bigr) \Pi_{DA}^\prime 
    \biggr), \nonumber \\[15pt]
\label{eq: STU_DZ_B}
& \alpha \Delta T_{\text{DZ-B}} \simeq 
    \frac{\Pi_{WW}}{m_W^2} - \frac{\Pi_{ZZ}}{M_Z^2}
    -2\xi \biggl( \frac{\Pi_{ZD}}{M_Z^2} + \Pi_{ZD}^\prime \biggr)
    + \hat{\varepsilon} s_W \Pi_{ZD}^\prime 
    + t_W\bigl(\hat{\varepsilon} s_W - 2\xi \bigr) \Pi_{DA}^\prime,  \\[15pt]
& \alpha \Delta U_{\text{DZ-B}} \simeq 
    4 s_W^2 \Bigl( 
        \Pi_{WW}^\prime  - c_W^2 \Pi_{ZZ}^\prime 
        - 2 s_W c_W \Pi_{ZA}^\prime - s_W^2 \Pi_{AA}^\prime 
        \Bigr). \nonumber 
\end{align}
Since Eqs.~(\ref{eq: seff_DP_STU})--(\ref{eq: MW_DP_STU}) also hold in RS-B, new physics effects from the oblique corrections are described by only the oblique parameters. 

The difference between the oblique parameters using RS-A and RS-B can be compensated by the relation between $\varepsilon$ and $\hat{\varepsilon}$:
\begin{align}
\label{eq: RS_conversion}
\hat{\varepsilon} \simeq \varepsilon\biggl\{ 
    & 1 - \frac{ 1 }{ 2\xi } \Pi_{ZD}^\prime 
    + \frac{ 1 }{ \xi (1-r^2) } \biggl( \frac{ \Pi_{ZD} }{ M_Z^2 } + \Pi_{ZD}^\prime \biggr)
    %
    %\nonumber \\[10pt]
    %
    %& 
    + \biggl( \frac{ 1 }{ \varepsilon s_W} - \frac{ 1 }{ 2\xi } \biggr)
        t_W \Pi_{DA}^\prime 
    \biggr\} + \frac{ 1 }{ s_W } \delta (\eta \varepsilon_Z)^{\overline{\mathrm{MS}}}.  
\end{align}
By substituting this relation to $S_\mathrm{tree}$, $T_\mathrm{tree}$, and $U_\mathrm{tree}$, we can reproduce the result in RS-A.

We comment in passing on the universality of the formulas for the $U$ parameter. 
In all models discussed above, the expressions for the $U$ parameter are identical. 
This is because $\Delta (\alpha U)$ in Eq.~(\ref{eq: STUparams_general_rep}) does not depend on $\Delta (\eta \varepsilon s_\theta)_\ast$, which is the factor that causes the difference among the models.  The $U$ parameter is insensitive to how $\varepsilon$ and $\varepsilon_Z^{}$ are defined.

%%%%%%%%%%%%%%%%%%%%%%%%%%%%%%%%%%%%%%%%%%%%%%%%%%
\subsection{An example: a dark doublet scalar field}
\label{subsec: example}
%%%%%%%%%%%%%%%%%%%%%%%%%%%%%%%%%%%%%%%%%%%%%%%%%%

Here, we consider a dark doublet scalar field $\phi_D$ as an example of new physics to evaluate the $S$, $T$, and $U$ parameters.
We assume that $\phi_D$ has $Y = 1/2$ and $Q_D = q$, and its VEV is zero. 
We do not consider the effects of the dark singlet, a new particle common to both types of models, because its contribution is of higher order. 
Also, we do not consider mixing between the neutral component of $\phi_D$ and the dark singlet for the sake of simplicity. 
Although the Higgs sector in the dark $Z$ models may include other additional scalar bosons, we also neglect their effects for the ease of a direct comparison with the dark photon models.

The dark doublet $\phi_D$ consists of a charged Higgs boson $H^\pm$ and two kinds of neutral Higgs bosons, a $CP$-even $H$ and a $CP$-odd $A$, and is thus parametrized as 
\begin{align}
\phi_D = \frac{ 1 }{ \sqrt{2}}
\begin{pmatrix}
\sqrt{2} H^\pm \\
H + i A\\
\end{pmatrix}.
\end{align}
We assume they have independent masses, $m_{H^\pm}$, $m_H$, and $m_A$, respectively, which are set to be
\begin{align}
m_H = 400~\mathrm{GeV}, \quad 
m_A = m_{H^\pm} = 200~\mathrm{GeV}. 
\end{align}
We have assumed the mass degeneracy between $A$ and $H^\pm$ to avoid a large one-loop contribution in the $T$ parameter, which is independent of the kinetic mixing. 
Since $\phi_D$ does not acquire a VEV, the one-loop corrections by $H^\pm$, $H$, and $A$ are gauge-invariant. 
Thus, the Ward--Takahashi identity $\Pi_{AA}(0) = \Pi_{ZA}(0) = \Pi_{DA}(0) = 0$ is automatically satisfied for the contributions from $\phi_D$. 

The current electroweak fit gives the results:
\begin{align}
\label{eq: EWPT_constraint}
S = 0.05\pm0.07, \quad T = 0.00 \pm 0.06,
\end{align}
by fixing $U=0$~\cite{PDG}.
In the following, we examine the parameter constraints from the $1\sigma$ bounds of the above in each dark $U(1)$ model and demonstrate how drastically the one-loop mixing effect changes the constraint on the model parameters.

We remark on two effects that could in principle modify the EWPO constraints in Eq.~(\ref{eq: EWPT_constraint}) for a light $Z_D$ boson with $M_D < M_Z$.
First, the neglected contribution from the direct $Z_D$ mediation, the third term of Eq.~(\ref{eq: neutral amplitudes}), can give a sizable effect on low-energy observables with $|q^2| \simeq M_D^2$~\cite{Davoudiasl:2012ag, Davoudiasl:2012qa, Davoudiasl:2014kua, Davoudiasl:2023cnc, Davoudiasl:2015bua}, whereas it is negligibly small at the $Z$ pole. 
It requires a full electroweak fit for each model to include this effect consistently.
Alternatively, a more conservative analysis is omitting low-energy observable data from the electroweak fit, which modifies the bound on the oblique parameters. 

Second, while setting $U=0$ is a good approximation when the new particles are sufficiently heavier than the $Z$ boson~\cite{Grinstein:1991cd}, the tree-level $U$ parameter can become comparable with the tree-level $S$ and $T$ parameters in the case of $M_D < M_Z$, and the fit result with $U \neq 0$ should be used instead of Eq.~(\ref{eq: EWPT_constraint}).\footnote{In contrast, even in the case of $M_D < M_Z$, the one-loop contribution to $U$ is much smaller than those to $S$ and $T$ as long as new particles in the loop diagrams are much heavier than the $Z$ boson.}
We have numerically confirmed that including $U\neq0$ does not significantly alter the constraint on the model parameters. 

Although these effects are relevant for deriving an accurate EWPO constraint for $M_D < M_Z$, we expect that they do not affect our main conclusion that one-loop corrections of $\mathcal{O}(\hbar \varepsilon)$ and $\mathcal{O}(\hbar \varepsilon_Z^{})$ can be significant. 
It would suffice to use the result with $U=0$ for the present study, which focuses on examining the impact of the novel one-loop mixing effects rather than deriving the precise constraints on the parameters.
Therefore, we employ Eq.~(\ref{eq: EWPT_constraint}) as the constraints in the following analysis.

%%%%%%%%%%%%%%%%%%%%%%%%%%%%%%%%%%%%%%%%%%%%%%%%%%
\subsubsection{Dark photon models}
%%%%%%%%%%%%%%%%%%%%%%%%%%%%%%%%%%%%%%%%%%%%%%%%%%

In the dark photon models, the one-loop corrections are given by
\begin{align}
\label{eq: STU_DP_IDM}
\left\{
\begin{array}{l}
\displaystyle{
	\Delta S_\mathrm{DP} = \frac{1}{2\pi} \biggl\{
	\biggl( 1 - 2a\xi \frac{ c_{2W}^{} }{ s_W^2 } \biggr) H(r_H,r_A)
		- \frac{a \xi }{s_W^2 M_Z^2} F(m_H,m_A) \biggr\}
	}, \\[15pt]
\displaystyle{
	\Delta T_\mathrm{DP} = \frac{\sqrt{2}G_F}{16\pi^2\alpha}
	\Bigl\{ F(m_H^{}, m_{H^\pm}^{})
		+ F(m_A^{}, m_{H^\pm}^{}) 
		- (1+4a\xi) F(m_H,m_A) \Bigr\}
	}, \\[15pt]
\displaystyle{
	\Delta U_\mathrm{DP} = \frac{1}{2\pi}
	\Bigl\{  H(1,r_H) + H(1,r_A) - (1+4a\xi) H(r_H, r_A) \Bigr\}
	}, \\
\end{array}
\right.
\end{align}
where $c_{2W} = \cos 2 \theta_W$, $a = q g_D^{}/g_Z^{}$, $r_H = m_H^2/m_{H^\pm}^2$, $r_A = m_A^2/m_{H^\pm}^2$, and $\xi = \varepsilon s_W/(1-r^2)$. 
The functions $F(m_1,m_2)$ and $H(x,y)$ are defined by
\begin{align}
& F(m_1, m_2) = \frac{m_1^2 + m_2^2}{2} - \frac{ m_1^2 m_2^2 }{ m_2^2 - m_1^2 } \log \frac{ m_2^2 }{ m_1^2 }, %\nonumber 
\\[5pt]
& H(x, y) = \frac{ 1 }{ 36(x-y)^3 }
	\Bigl\{ (y-x)(5x^2 - 22xy + 5y^2) 
		%
%		\nonumber \\
		%
%		& \hspace{120pt} 
        + 6x^2 (x-3y) \log x
		+ 6y^2(3x-y)\log y \Bigr\}. %\nonumber 
\end{align}
In the limit of $\varepsilon \to 0$ or $q \to 0$, 
the one-loop corrections coincide with those in the inert doublet model (IDM)~\cite{Barbieri:2006dq, Merchand:2019bod}.   
The terms proportional to $a \xi$ are novel one-loop contributions caused by the mixing.
They strongly depend on the mass difference between $H$ and $A$.

The parameter $\Delta T_\mathrm{DP}$ does not vanish even if $m_{H^\pm} = m_H$ or $m_{H^\pm} = m_A$ as long as $\xi \neq 0$ although these conditions lead to $T=0$ in the IDM. 
This is because the mixing violates the (twisted) custodial symmetry~\cite{Sikivie:1980hm, Pomarol:1993mu, Gerard:2007kn, deVisscher:2009zb}.

\begin{figure}[t]
\begin{center}
\includegraphics[width=0.6\textwidth]{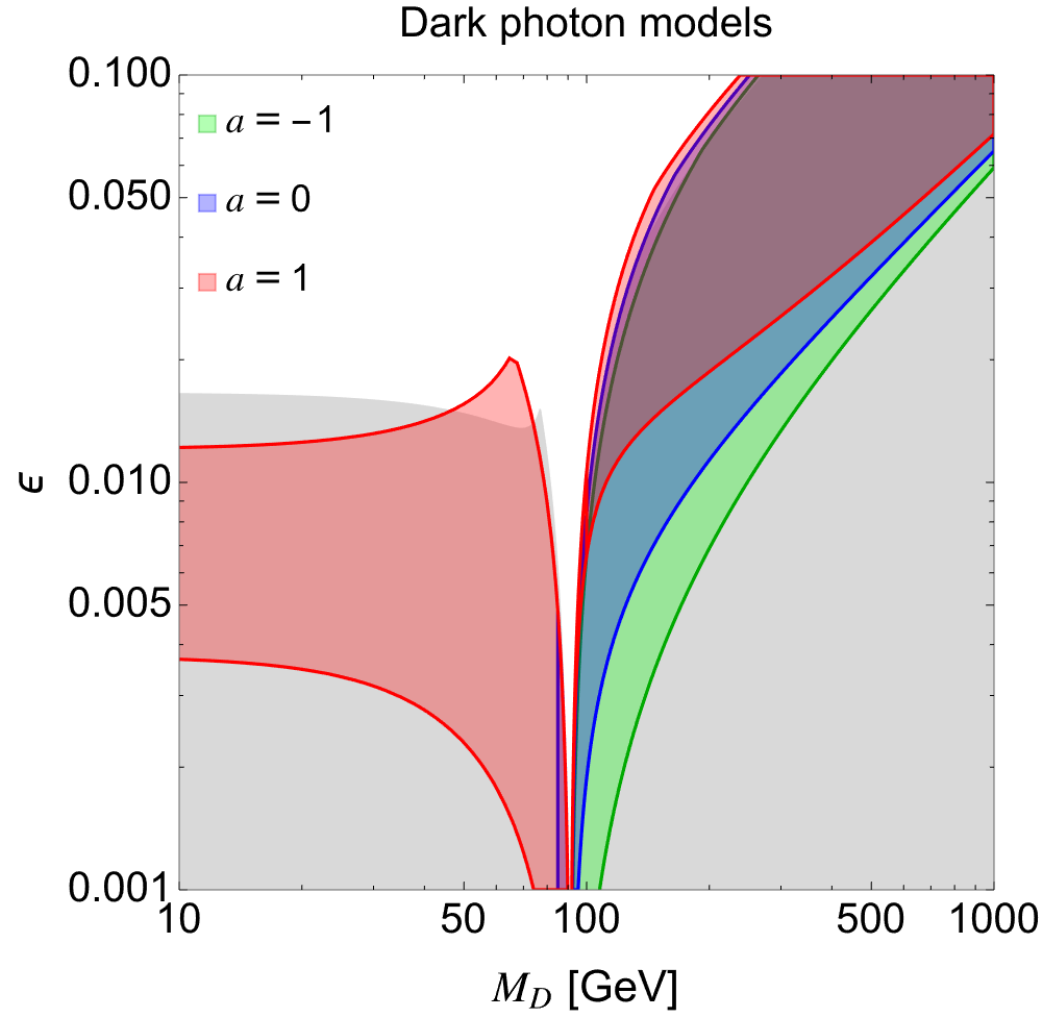}
\caption{The EWPO constraints in the $M_D$-$\varepsilon$ plane. The gray region is the allowed region at tree level. The red, blue, and green regions are those at one-loop level with $a = 1$, $0$, and $-1$, respectively. The model parameters are fixed as $m_H = 400~\mathrm{GeV}$, $m_A = m_{H^\pm}^{} = 200~\mathrm{GeV}$.}
\label{fig: DP}
\end{center}
\end{figure}

In Fig.~\ref{fig: DP}, we show regions constrained by the current electroweak precision measurements in the $M_D$-$\varepsilon$ plane. 
The gray region is the allowed region at tree level using Eq.~(\ref{eq: STU_tree}). 
The red, blue, and green regions represent the allowed regions at one-loop level, which are given by the sum of Eq.~(\ref{eq: STU_DP_IDM}) and their tree-level formulas, with $a=1$, $0$, and $-1$. 

We assume a positive $\varepsilon$ in making Fig.~\ref{fig: DP}. 
The result for a negative $\varepsilon$ is equivalent to switching the sign of $a$; for example, the result for $(\varepsilon, a) = (-10^{-2}, 1)$ is the same that for $(\varepsilon, a) = (10^{-2}, -1)$ because all the terms linear in $\varepsilon$ are proportional to $a$ in the oblique parameters.

One can see that the one-loop effects significantly change the allowed region from the tree-level result. 
The blue region ($a=0$) correspond to the IDM, as explained above. 
In this case, most of the allowed region is in the domain $M_D > M_Z$, which makes $\xi$ negative. 
It is because a large one-loop correction in $\Delta S_\mathrm{DP}|_{a=0}$ is positive, independently of $\varepsilon$ and $M_D$. 
To cancel it, $S_\mathrm{tree}$ needs to be negative, which leads to $r = M_D/M_Z > c_W \simeq 0.9$. 
It is worth noting that smaller values of $\varepsilon$ can be excluded unless $M_D \simeq M_Z$ because the size of $S_\mathrm{tree}$ is not sufficiently large to reduce the $\Delta S_\mathrm{DP}$ in such regions.   

In the case of $a=-1$ (the green region), the allowed region exists only when $M_D > M_Z$ because the terms proportional to $a \xi$ increase $\Delta S_\mathrm{DP}$ if $\xi >0$. 
Thus, a positive $\xi$ makes the situation more constraining than the case of $a=0$. 
If $\xi < 0$, the one-loop correction is reduced and more allowed region appears, which is different from that for $a=0$.

If $a=1$ (the red region), the allowed region is significantly different from those for the other cases. 
In this case, a positive $\xi$ makes $\Delta S_\mathrm{DP}$ smaller. 
It enables the models to avoid the constraint in the region $M_D < M_Z$. 

As seen above, the one-loop mixing effect is important to evaluate the EWPO constraint. 
The result can be quite different from both the tree-level result and that in the IDM.

%%%%%%%%%%%%%%%%%%%%%%%%%%%%%%%%%%%%%%%%%%%%%%%%%%
\subsubsection{Dark $Z$ models using RS-A}
%%%%%%%%%%%%%%%%%%%%%%%%%%%%%%%%%%%%%%%%%%%%%%%%%%

In the dark $Z$ models using RS-A, the one-loop contributions are given by
\begin{align}
\left\{
\begin{array}{l}
\displaystyle{
    \Delta S_{\text{DZ-A}} = 
    \frac{1}{2\pi} \Biggl[
	\biggl\{ 1 - \frac{2a}{s_W^2}\Bigl(\xi c_{2W} - \xi_Z \Bigr)
        \biggr\} H(r_H,r_A)
	- \frac{a(\xi - \xi_Z)}{s_W^2M_Z^2} F(m_H,m_A)
    \Biggr]
    }, \\[15pt]
\displaystyle{
	\Delta T_{\text{DZ-A}} = \frac{\sqrt{2}G_F}{16\pi^2\alpha}
	\Bigl\{ F(m_H^{}, m_{H^\pm}^{})
		+ F(m_A^{}, m_{H^\pm}^{})} \\
            \hspace{100pt}
		\displaystyle{
            - (1+4a\xi-4a\xi_Z) F(m_H,m_A) 
            + 8a \xi_Z M_Z^2 H(r_H,r_A) \Bigr\}
	}, \\[15pt]
\displaystyle{
    \Delta U_{\text{DZ-A}} = 
    \frac{1}{2\pi}
	\Bigl\{  H(1,r_H) + H(1,r_A) - (1+4a\xi) H(r_H, r_A) \Bigr\}
    }, \\
\end{array}
\right.
\end{align}
where 
\begin{align}
\xi = \frac{\varepsilon_Z^{} + \varepsilon s_W }{ 1-r^2 }, \quad 
\xi_Z = \frac{\varepsilon_Z^{}  }{ 1-r^2 }.
\end{align}
To derive the $S$ and $T$ parameters, we have used 
\begin{align}
\delta (\eta \varepsilon_Z^{})^{\overline{\mathrm{MS}}}
= \frac{ a \alpha }{ 12\pi \xi c_W^2 } \frac{ \varepsilon_Z^{} }{ 1-r^2 } \Xi, 
\end{align}
where $\Xi$ is the divergent part in the $\overline{\mathrm{MS}}$ scheme (see Appendix~\ref{app: divergence_cancellation}).
In the limit of $\varepsilon_Z^{} \to 0$, these coincide with those in the dark photon models. 
Therefore, in the dark $Z$ models using RS-A, we can readily take the dark photon limit including one-loop corrections.

\begin{figure}[t]
\begin{center}
\subfigure[\ $M_D = 10~\mathrm{GeV}$, $\varepsilon_Z^{}>0$]{
    \includegraphics[width=0.48\textwidth]{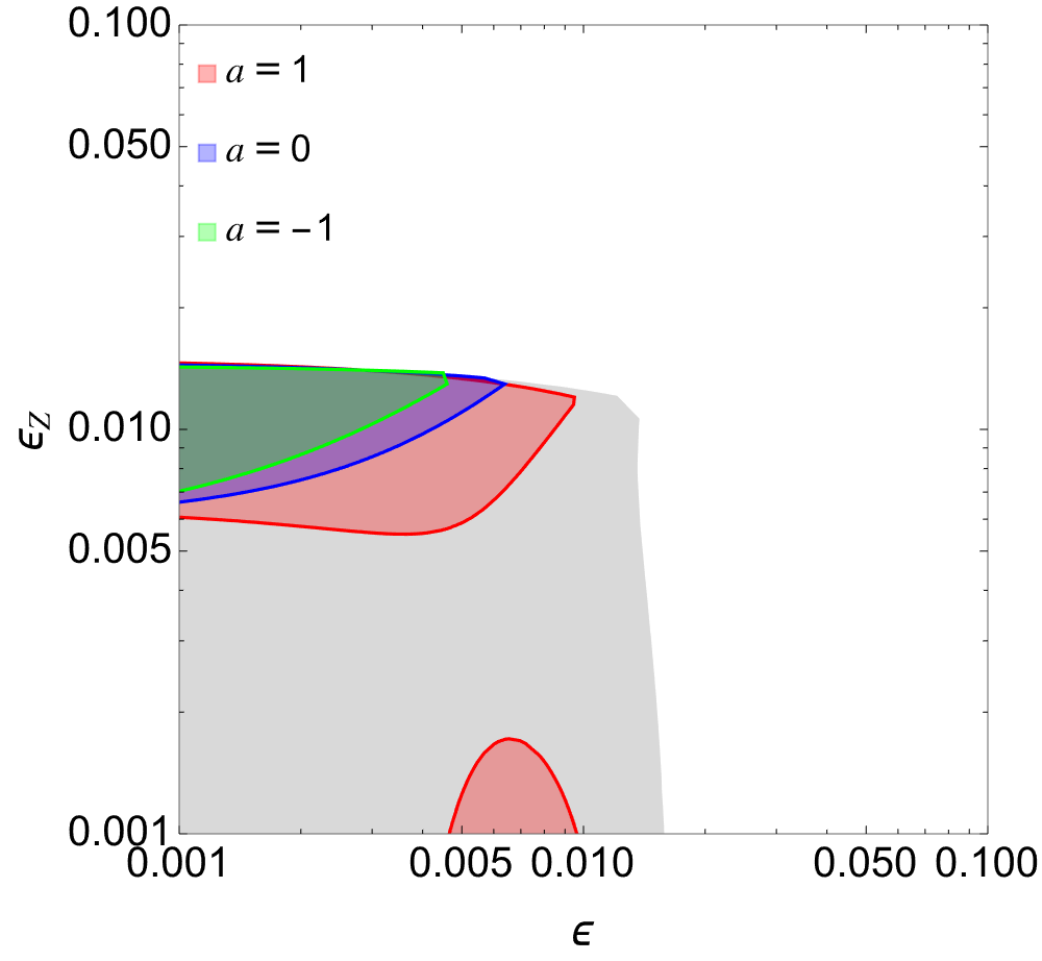}
    \label{fig: DZ_A_Low_SS}
}
\subfigure[\ $M_D = 300~\mathrm{GeV}$, $\varepsilon_Z^{}>0$]{
    \includegraphics[width=0.48\textwidth]{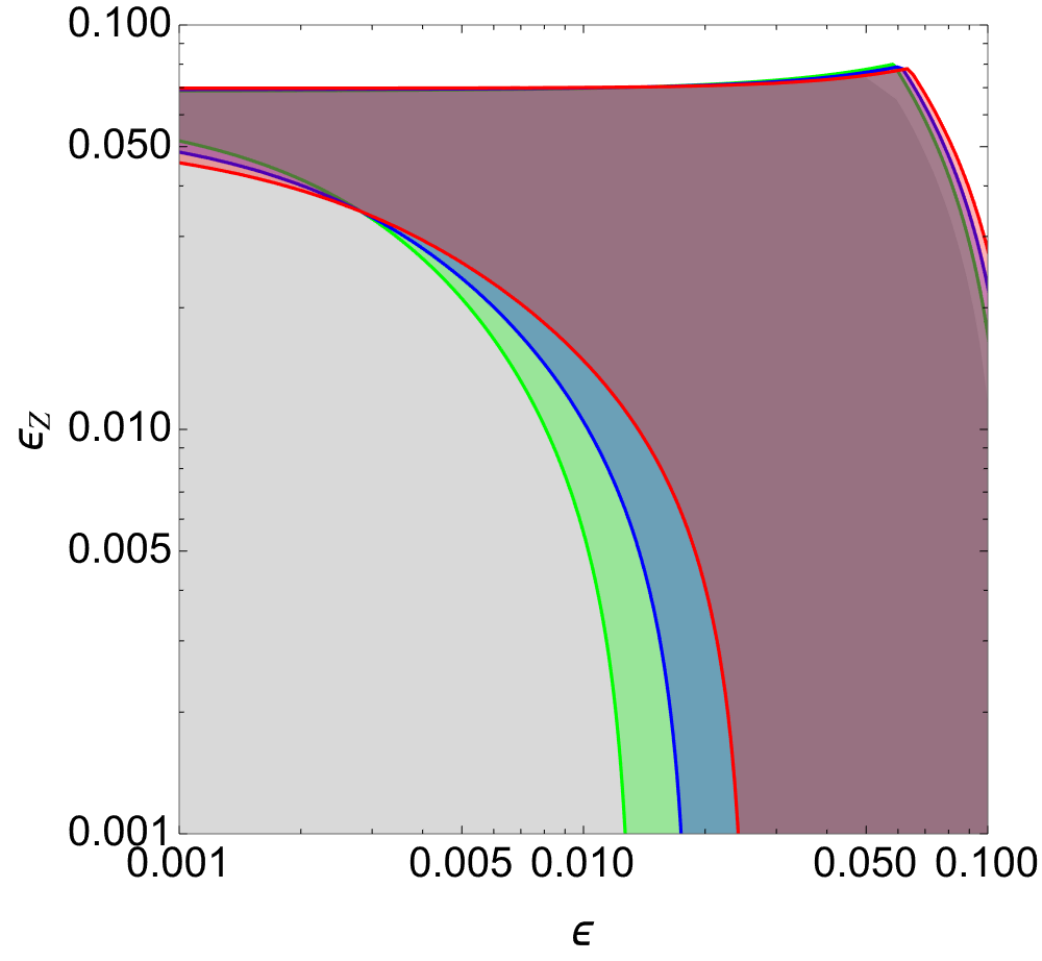}
    \label{fig: DZ_A_High_SS}
}
\subfigure[\ $M_D = 10~\mathrm{GeV}$, $\varepsilon_Z^{}<0$]{
    \includegraphics[width=0.48\textwidth]{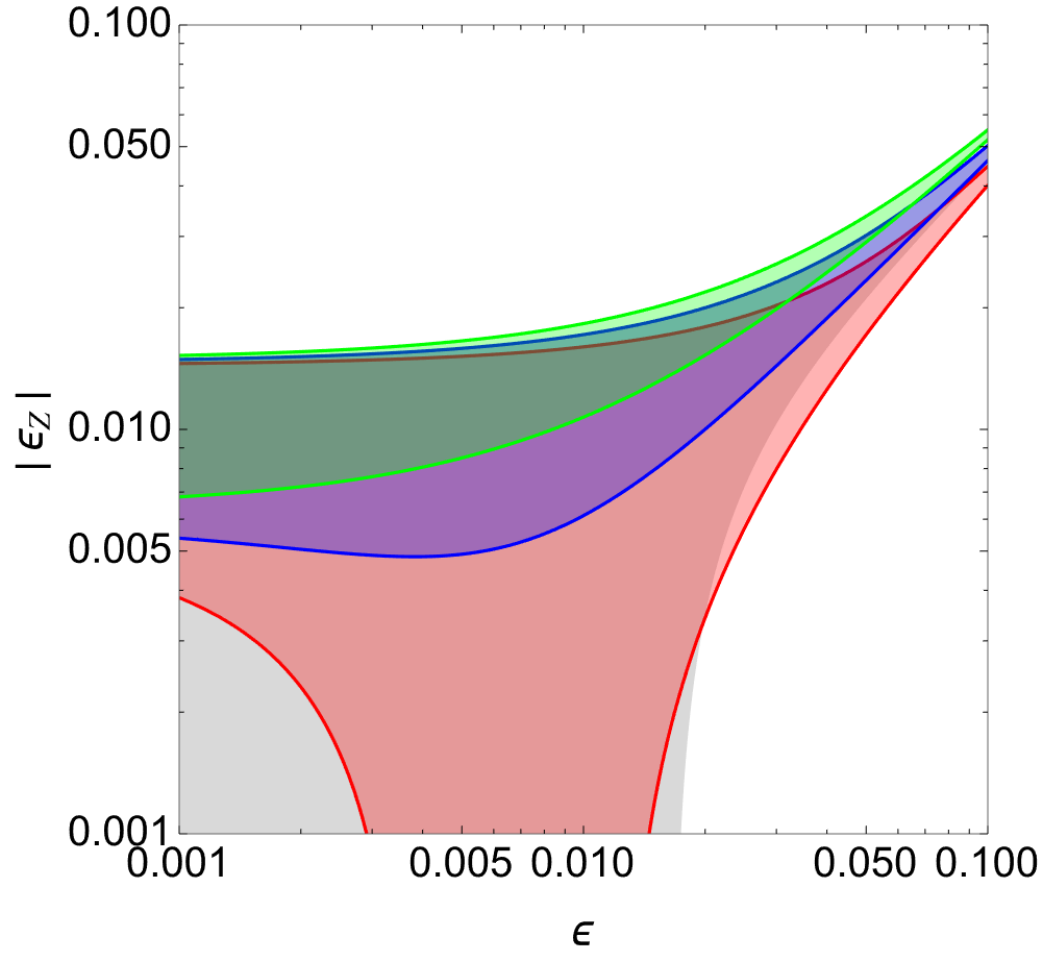}
    \label{fig: DZ_A_Low_OS}
}
\subfigure[\ $M_D = 300~\mathrm{GeV}$, $\varepsilon_Z^{}<0$]{
    \includegraphics[width=0.48\textwidth]{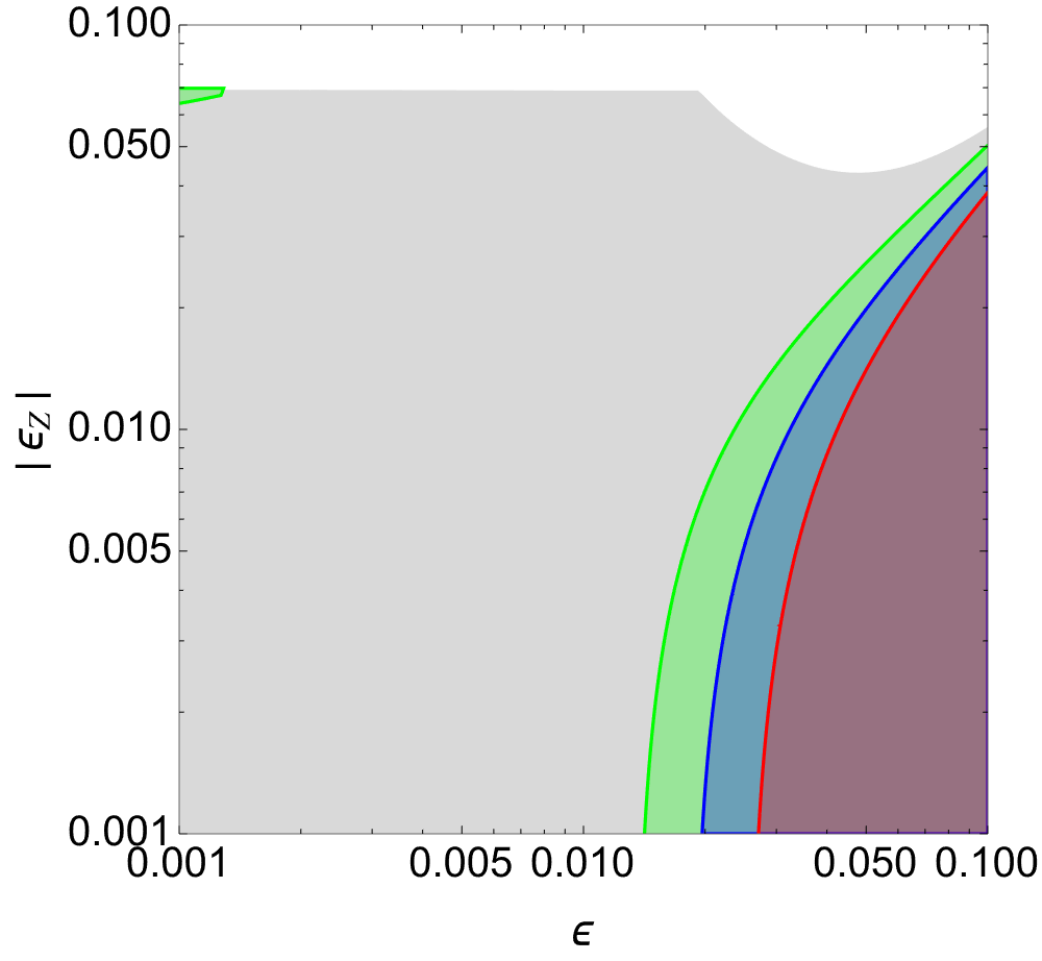}
    \label{fig: DZ_A_High_OS}
}
\caption{Constraints in the $\varepsilon$-$\varepsilon_Z^{}$ plane in the dark $Z$ models using RS-A, for different choices of $M_D$ and $\varepsilon_Z$. The color scheme is the same as in Fig.~\ref{fig: DP}.}
\label{fig: DZ_A}
\end{center}
\end{figure}

In Fig.~\ref{fig: DZ_A}, we show the allowed region in the plane of $\varepsilon$ and $\varepsilon^{}_Z$ due to the constraints from the $S$ and $T$ parameters~(\ref{eq: EWPT_constraint}). 
The meaning of the colors and the masses of the dark doublet scalars are the same as in Fig.~\ref{fig: DP}. 
The mass of $Z_D$ is set to be $10~\mathrm{GeV}$ in Fig.~\ref{fig: DZ_A_Low_SS} and Fig.~\ref{fig: DZ_A_Low_OS} and to be $300~\mathrm{GeV}$ in Fig.~\ref{fig: DZ_A_High_SS} and Fig.~\ref{fig: DZ_A_High_OS}. In all the figures, $\varepsilon$ is fixed to be positive; on the other hand, $\varepsilon_Z^{}$ is positive in the upper figures and negative in the lower figures. 
The constraints on negative $\varepsilon$ are given by changing the sign of $a$ in the corresponding figure of Fig.~\ref{fig: DZ_A}, as in the dark photon models.

In Figs.~\ref{fig: DZ_A_Low_SS} and~\ref{fig: DZ_A_Low_OS} with $M_D = 10~\mathrm{GeV}$, 
there are allowed regions for small $\varepsilon^{}_Z$ only in the case of $a=1$. 
This is consistent with Fig.~\ref{fig: DP}, which corresponds to the dark photon limit of the dark $Z$ model with RS-A. 
The difference between Figs.~\ref{fig: DZ_A_Low_SS} and~\ref{fig: DZ_A_Low_OS} is caused by the fact that the sign of the mass mixing changes the interference among the one-loop contributions.

On the other hand, in Figs.~\ref{fig: DZ_A_High_SS} and~\ref{fig: DZ_A_High_OS}, there are allowed regions for small $\varepsilon_Z^{}$ in all the cases. 
This can also be understood as the behavior approaching the dark photon limit in Fig.~\ref{fig: DP}. 
The difference between these two cases is mainly in the behavior for small $\varepsilon$. 
In Fig.~\ref{fig: DZ_A_High_SS}, the allowed regions for smaller $\varepsilon$ are connected with those for relatively large $\varepsilon$. 
However, they are disconnected in Fig.~\ref{fig: DZ_A_High_OS} because of the interference of the one-loop contributions. 
Although there are narrow allowed regions for small $\varepsilon$ and negative $\varepsilon_Z^{}$, the regions for $a=1$ and $0$ are out of the range of the figure, and part of that for $a=-1$ is shown around $|\varepsilon_Z^{}| \simeq 0.07$ in Fig.~\ref{fig: DZ_A_High_OS}.

Consequently, in all the cases, the allowed regions are significantly changed and restricted by the one-loop corrections and depend on the value of $a$. 
The one-loop mixing effect plays an important role in examining the constraints, in particular, for the lighter $Z_D$ bosons.

%%%%%%%%%%%%%%%%%%%%%%%%%%%%%%%%%%%%%%%%%%%%%%%%%%
\subsubsection{Dark $Z$ models using RS-B}
%%%%%%%%%%%%%%%%%%%%%%%%%%%%%%%%%%%%%%%%%%%%%%%%%%

When we employ RS-B in the dark $Z$ models, the one-loop corrections are given by
\begin{align}
\left\{
\begin{array}{l}
\displaystyle{
    \Delta S_{\text{DZ-B}} = 
    \frac{ 1 }{ 2\pi } \biggl( 1 + 4a\xi - \frac{a\hat{\varepsilon}}{s_W} \biggr)
    H(r_H, r_A)
    }, \\[15pt]
\displaystyle{
    \Delta T_{\text{DZ-B}} = \frac{\sqrt{2}G_F}{16\pi^2\alpha}
    \Bigl\{ F(m_H^{}, m_{H^\pm}^{}) + F(m_A^{}, m_{H^\pm}^{})
        - F(m_H^{},m_A^{}) } \\
        \hspace{100pt} 
        \displaystyle{
        + 4a (2\xi - \hat{\varepsilon}s_W)M_Z^2 H(r_H,r_A)
    \Bigr\}
    }, \\[15pt]
\displaystyle{
    \Delta U_{\text{DZ-B}} = \frac{1}{2\pi}
    \Bigl\{  H(1,r_H) + H(1,r_A) - (1+4a\xi) H(r_H, r_A) \Bigr\}
    }, \\
\end{array}
\right.
\end{align}
where $\xi = (\hat{\varepsilon}_Z + \hat{\varepsilon} s_W)/(1-r^2)$. 
The results in the standard EW gauge theory are reproduced in the no-mixing limit or $q \to 0$. 
However, even in the limit of $\hat{\varepsilon}_Z^{} \to 0$, the oblique parameters are different from those in the dark photon models. 

\begin{figure}[t]
\begin{center}
\subfigure[\ $M_D = 10~\mathrm{GeV}$, $\hat{\varepsilon}_Z>0$]{
    \includegraphics[width=0.48\textwidth]{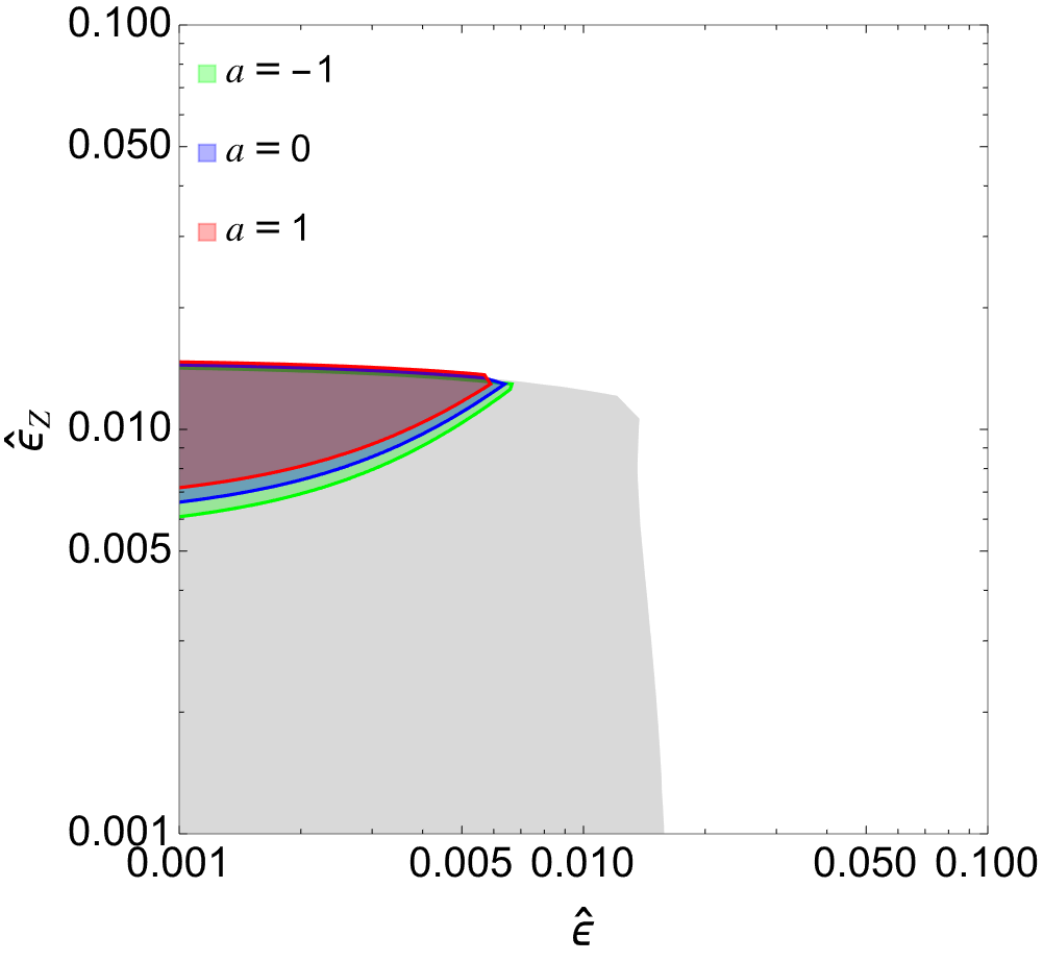}
    \label{fig: DZ_B_Low_SS}
}
\subfigure[\ $M_D = 300~\mathrm{GeV}$, $\hat{\varepsilon}_Z>0$]{
    \includegraphics[width=0.48\textwidth]{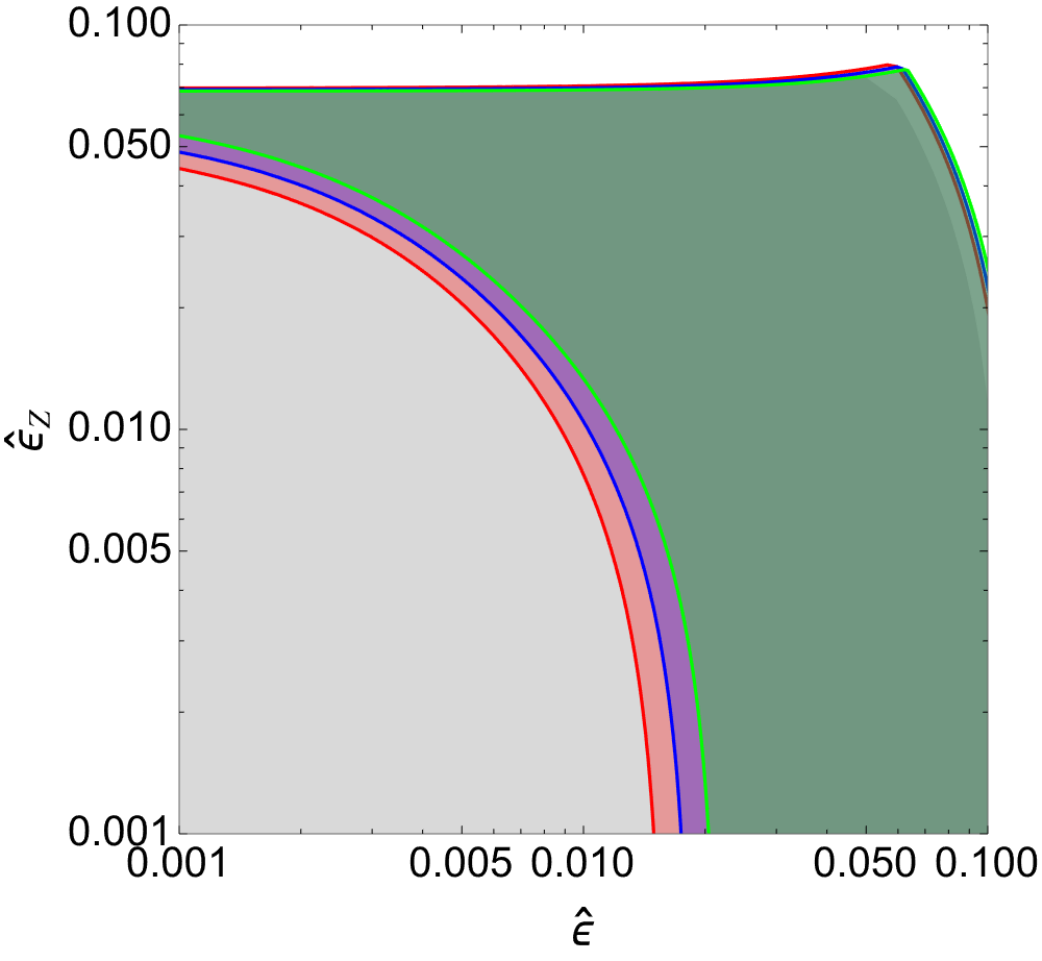}
    \label{fig: DZ_B_High_SS}
}
\subfigure[\ $M_D = 10~\mathrm{GeV}$, $\hat{\varepsilon}_Z<0$]{
    \includegraphics[width=0.48\textwidth]{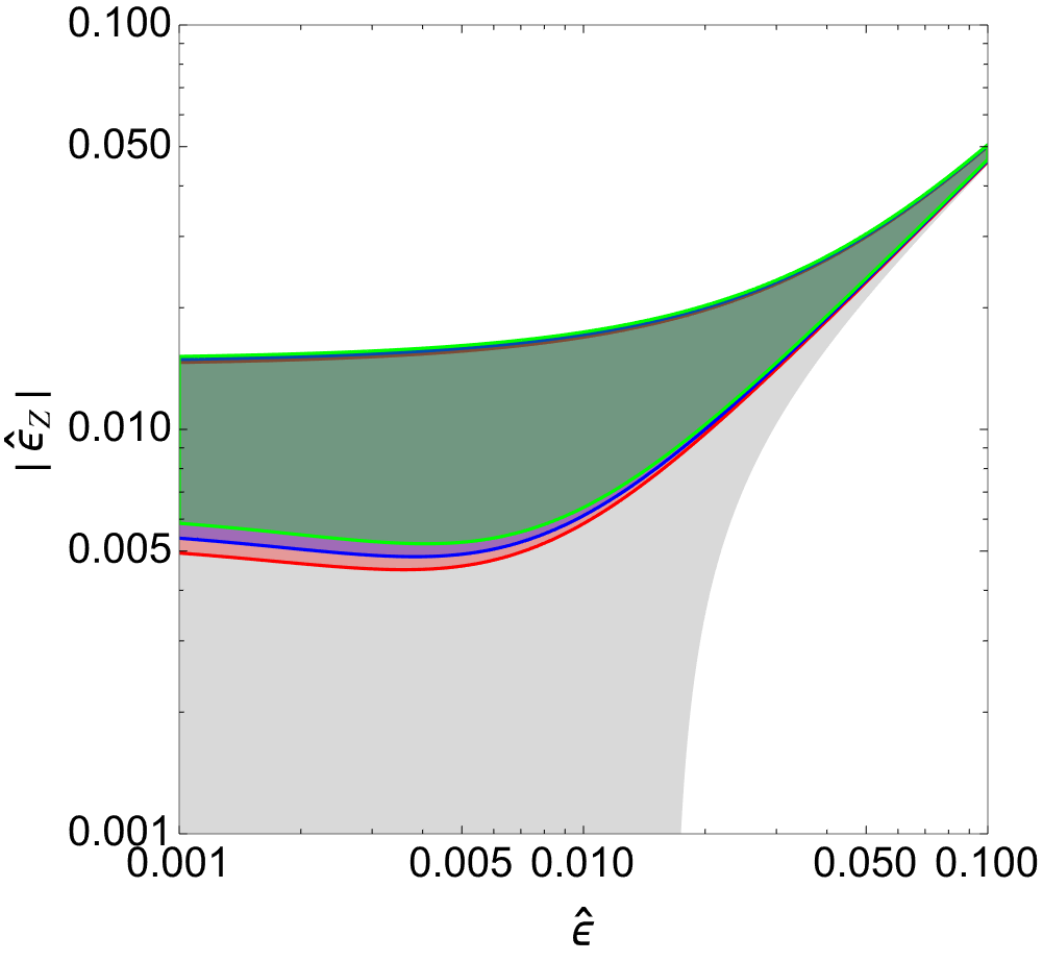}
    \label{fig: DZ_B_Low_OS}
}
\subfigure[\ $M_D = 300~\mathrm{GeV}$, $\hat{\varepsilon}_Z<0$]{
    \includegraphics[width=0.48\textwidth]{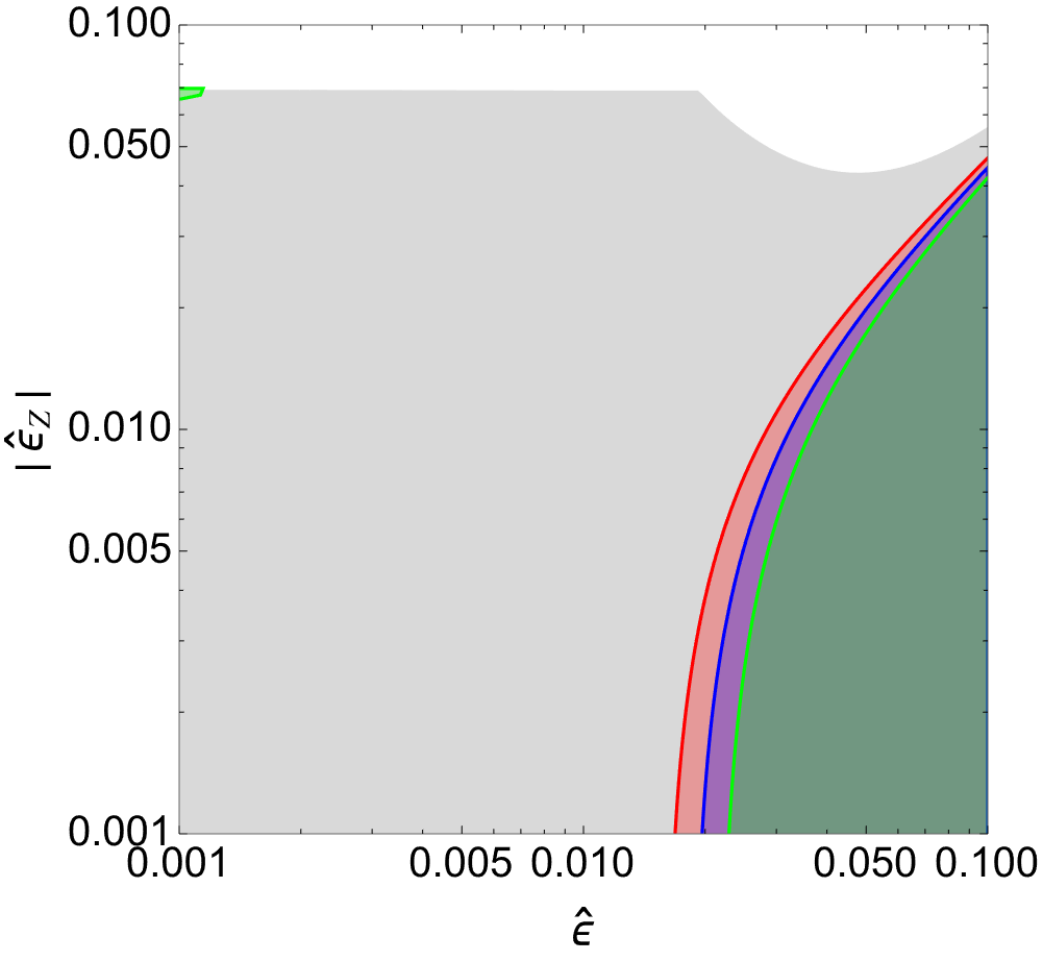}
    \label{fig: DZ_B_High_OS}
}
\caption{Constraints in the $\hat{\varepsilon}$-$\hat{\varepsilon}_Z$ planes in the dark $Z$ models using RS-B. The color scheme is same as in Fig.~\ref{fig: DP}.}
\label{fig: DZ_B}
\end{center}
\end{figure}

In Fig.~\ref{fig: DZ_B}, we show the constraints in the plane of $\hat{\varepsilon}$ and $\hat{\varepsilon}_Z$ using the same input parameters as in Fig.~\ref{fig: DZ_A}. 
We observe that the dependence on $a$ is weaker than that using RS-A although the one-loop effect still drastically changes the allowed regions in all the cases. 
In other words, the perturbative expansion by the mixing parameters would be more stable in RS-B. 
This is because the one-loop correction in $(\hat{\eta}\hat{\varepsilon}s_\theta)_\ast$ is absorbed into the definition of $\hat{\varepsilon}$. 
However, as mentioned above, it is difficult in this scheme to compare the results with those in the dark photon models at one-loop level.

%%%%%%%%%%%%%%%%%%%%%%%%%%%%%%%%%%%%%%%%%%%%%%%%%%
\section{Conclusions}
\label{sec: conclusions}
%%%%%%%%%%%%%%%%%%%%%%%%%%%%%%%%%%%%%%%%%%%%%%%%%%

In this paper, we have considered the extension of the oblique parameters to the dark $U(1)$ models. 
We have thoroughly investigated the oblique corrections in the four-fermion process by solving the Schwinger--Dyson equation for the gauge boson propagators. 
We have defined the running parameters at the one-loop level. 
The mixing effect has been included up to the quadratic and linear order terms at tree and one-loop levels, respectively. 
We have considered two classes of models: the dark photon models and the dark $Z$ models. 
For the latter, we have employed two renormalization schemes (RSs): RS-A and RS-B.

When the new physics scale is much higher than the $Z$ boson mass, we have shown that the oblique corrections can be described by the $S$, $T$, and $U$ oblique parameters except for the effects mediated by the $Z_D$ boson. 
As an explicit example, we have considered the dark isospin doublet $\phi_D$.
We have shown the current electroweak precision observables (EWPO) constraints for each class of models and have seen that the allowed regions in the mixing parameters would drastically change by the dark charge of $\phi_D$ except for the dark $Z$ models with RS-B. 
In the latter, although the perturbation by the mixing parameters is more stable, it is difficult to compare the result with the dark photon models because we cannot take the dark photon limit.

In the future GIGA-Z experiment, the errors at $1\sigma$ level are expected to be improved to $0.02$ for the $S$ and $T$ parameters ({\it i.e.}, shrinking by roughly a factor of 3 from the current errors) with fixing $U=0$~\cite{Erler:2000jg}. 
In such a case, the novel mixing effects at the one-loop level in our results will become more significant to examine the EWPO constraint.

In our analysis of the EWPO constraints, we have neglected the effects of the $Z_D$ boson mediation, the third term in Eq.~(\ref{eq: neutral amplitudes}). 
As excused in Sec.~\ref{subsec: example}, this term can affect the observables at the scale $|q^2| \simeq M_D^2$, such as the weak mixing angle measurements at low energies. 
To use electroweak observables off the $Z$ pole, we need to factorize this effect in addition to the result in this paper. 
This would be a worthwhile issue for future research. 
In addition, we have not discussed in detail how to construct the gauge-independent two-point functions for the gauge bosons by using the pinch technique in the dark $U(1)$ models. 
This is crucial in, for example, the simplest dark $Z$ model introduced in Appendix~\ref{app: darkZ_model}, where $U(1)_D$ is broken by the VEVs of dark doublet and singlet scalars, because the Nambu-Goldstone mode for $U(1)_D$ breaking contributes to the 1PI diagrams, which is not the SM contribution. 
The full gauge-invariant result in this model will be discussed in a separate analysis~\cite{CE}.

%%%%%%%%%%%%%%%%%%%%%%%%%%%%%%%%%%%%%%%%%%%%%%%%%%
\section*{Acknowledgments}
This work was supported by the National Science and Technology Council under Grant Nos.~NSTC-111-2112-M002-018-MY3, NSTC-113-2811-M-002-043, and NSTC-114-2112-M-002-020-MY3.

\appendix

%%%%%%%%%%%%%%%%%%%%%%%%%%%%%%%%%%%%%%%%%%%%%%%%%%
\section{An example of dark $Z$ models}
\label{app: darkZ_model}
%%%%%%%%%%%%%%%%%%%%%%%%%%%%%%%%%%%%%%%%%%%%%%%%%%

In this appendix, we introduce the simplest example of the dark $Z$ model proposed in Ref.~\cite{Davoudiasl:2012ag}, where both the kinetic mixing $\varepsilon$ and the mass mixing $\varepsilon_Z^{}$ exist in the mass matrix of the neutral gauge bosons. 
All of formulas in this section are at tree level. 
Thus, we do not use a subscript $0$ to represent bare parameters.

The Higgs sector of the model consists of three kinds of scalar fields: an isospin doublet $\phi_1$ with $Y=1/2$ and $Q_D=0$, an isospin doublet $\phi_2$ with $Y = 1/2$ and $Q_D=1$, and a isospin singlet $S$ with $Y = 0$ and $Q_D = q_s$. 
They acquire VEVs as 
\begin{align}
    \left<\phi_i\right> = \frac{ 1 }{ \sqrt{2} }(0, v_i)^\mathrm{T}, \quad 
    \left< S \right> = \frac{ v_S }{ \sqrt{2} }, 
\end{align}
for $i=1$, 2. 
The electroweak symmetry is broken by $v_1$ and $v_2$, while the dark symmetry is broken by $v_2$ and $v_S$. 
We define the electroweak VEV $v = \sqrt{v_1^2 + v_2^2} \simeq 246~\mathrm{GeV}$ and the angle $\beta$ as 
\begin{align}
\tan \beta = \frac{ v_2 }{ v_1 }. 
\end{align}

The mass matrix for $\tilde{Z}$ and $\tilde{Z}_D$, which are defined as explained in Sec.~\ref{sec: models}, is given by
\begin{align}
M_V^2 = 
\begin{pmatrix}
\tilde{m}_Z^2 & -\tilde{m}_Z^2 \eta (\varepsilon_Z + \varepsilon s_\theta ) \\
-\tilde{m}_Z^2 \eta (\varepsilon_Z + \varepsilon s_\theta ) & \tilde{m}_{D}^2 + \tilde{m}_Z^2 \eta^2  \varepsilon^2 s_\theta^2  \\
\end{pmatrix},
\end{align}
where
\begin{align}
\tilde{m}_Z^2 = \frac{ g_Z^2 v^2 }{ 4 }, \quad 
\tilde{m}_D^2 = \eta^2 g_D^2 (v_2^2 + q_S^2 v_S^2 ) + 2 \tilde{m}_Z^2\eta^2\varepsilon_Z \varepsilon s_\theta, \quad 
\varepsilon_Z = 2 \frac{ g_D }{ g_Z } \sin^2 \beta. 
\end{align}
Thus, this model has an additional mixing parameter, the mass mixing $\varepsilon_Z^{}$, which is independent of the kinetic mixing $\varepsilon$.

%%%%%%%%%%%%%%%%%%%%%%%%%%%%%%%%%%%%%%%%%%%%%%%%%%
\section{Effects of the absorptive parts of oblique corrections}
\label{app: absorptive_part}
%%%%%%%%%%%%%%%%%%%%%%%%%%%%%%%%%%%%%%%%%%%%%%%%%%

In this appendix, we discuss the case that the two-point functions have the imaginary part, {\it i.e.}, the absorptive part, and show the full one-loop expressions for the four-fermion amplitudes. 
In this case, the running parameters discussed in Sec.~\ref{sec: renormalization} are defined using only their real parts, {\it i.e.}, the dissipative parts, because the original parameters are real.

The absorptive parts arise if the external momentum exceeds the threshold for the internal particles to go on shell and are finite at the one-loop level. 
They are not renormalized and can induce a new type of interactions among the external fermions.

Let $\Pi_{VV^\prime}^\mathrm{Im}$ be the imaginary part of $\Pi_{VV^\prime}$. 
The four-fermion amplitudes including corrections by the absorptive parts are then given by
\begin{align}
\mathcal{M}_\mathrm{NC} = 
	& \frac{ e_\ast^2 Q Q^\prime }{ q^2\bigl(1 - i \tilde{\Pi}_{AA}^\mathrm{Im} \bigr) }
	+ V_{Z\ast}^f \frac{ 1 }{ q^2 - M_{Z\ast}^2 + i \sqrt{q^2} \Gamma_{Z\ast}(q^2) } V_{Z\ast}^{f^\prime} 
    \nonumber \\[5pt]
	& + V_{D\ast}^f \frac{ 1 }{ q^2 - M_{D\ast}^2 + i \sqrt{q^2} \Gamma_{D\ast}(q^2)  } V_{D\ast}^{f^\prime}, \\[10pt]
\mathcal{M}_\mathrm{CC} = 
    & \frac{ e_\ast^2 }{ 2 s_{\theta\ast}^2 }
    I_+ \frac{ Z_{W\ast} }{ q^2 - M_{W\ast}^2 + i\sqrt{q^2} \Gamma_{W\ast}(q^2) } I_-, 
\end{align}
where $\Gamma_{V\ast}$ ($V=Z$, $D$, and $W$) are defined by 
\begin{align}
\sqrt{q^2} \Gamma_{V\ast}(q^2) = - \Pi_{VV}^\mathrm{Im}(q^2). 
\end{align}
We note that $\Gamma_{V\ast}(q^2)$ is zero for space-like momentum $q^2<0$. Thus, the left-hand side of this equation is always real. 
Due to the optical theorem, $\Gamma_{V\ast}$ is equivalent to the decay width $\Gamma_V$ on the mass shell; {\it i.e.},
\begin{align}
\Gamma_{V\ast}(M_V^2) = \Gamma_V.
\end{align}

The running vertex factors $V_{Z\ast}^f$ and $V_{D\ast}^f$ are given by
\begin{align}
& V_{Z\ast}^f \simeq Z_{Z\ast}^{1/2} \frac{ e_\ast }{ s_{\theta\ast}^{} c_{\theta\ast}^{} }
	\biggl\{
	  \Bigl( 
            c_{\xi\ast}^{} 
            - is_\xi \zeta^\mathrm{Im}
            - is_\xi t_\theta \tilde{\Pi}_{\tilde{D}A}^\mathrm{Im}
        \Bigr)
        \Bigl( I_3 - s_{\theta\ast}^2 Q 
            + i s_\theta c_\theta  \tilde{\Pi}_{\tilde{Z}A}^\mathrm{Im} Q
        \Bigr) \nonumber \\
	& \hspace{100pt} 
        - \Bigl( s_{\xi\ast}^{} 
            + i c_\xi \zeta^\mathrm{Im} \Bigr)
        \Bigl( \eta_\ast \varepsilon_\ast s_{\theta\ast}
            + i t_\theta \tilde{\Pi}_{\tilde{D}A}^\mathrm{Im}
        \Bigr) (Q-I_3)
	\biggr\}, \\[10pt]
& V_{D\ast}^f \simeq Z_{D\ast}^{1/2} \frac{ e_\ast }{ s_{\theta\ast}^{} c_{\theta\ast}^{} }
    \biggl\{ 
        \Bigl( s_{\xi\ast} + i c_\xi \zeta^\mathrm{Im}
            + i c_\xi t_\theta \tilde{\Pi}_{\tilde{D}A}^\mathrm{Im} 
        \Bigr)
        \Bigl( 
            I_3 - s_\ast^2 Q + i s_\theta c_\theta \tilde{\Pi}_{\tilde{Z}A}^\mathrm{Im} Q
        \Bigr)
        \nonumber \\
        & \hspace{100pt} 
        - \Bigl( c_{\xi\ast}
            - i s_\xi \zeta^\mathrm{Im}
        \Bigr)
        \Bigl( \eta_\ast \varepsilon_\ast s_{\theta\ast}
            + i t_\theta \tilde{\Pi}_{\tilde{D}A}^\mathrm{Im}
        \Bigr)
        (Q-I_3)
    \biggr\}, 
\end{align}
where $\zeta^\mathrm{Im}$ is the imaginary part of the 
momentum-dependent angle $\zeta(q^2)$,
\begin{align}
\zeta^\mathrm{Im}(q^2) \simeq - \frac{ \Pi_{ZD}^\mathrm{Im} (q^2) }{ M_Z^2 - M_D^2 }.
\end{align}
We note that even when the two-point functions have an absorptive part, the definition of $\zeta(q^2)$ stays the same as in Eq.~(\ref{eq: zeta_def}).

%%%%%%%%%%%%%%%%%%%%%%%%%%%%%%%%%%%%%%%%%%%%%%%%%%
\section{Cancellation of the divergence in the running parameters}
\label{app: divergence_cancellation}
%%%%%%%%%%%%%%%%%%%%%%%%%%%%%%%%%%%%%%%%%%%%%%%%%%

In this appendix, we discuss the cancellation of divergences in the running parameters discussed in Sec.~\ref{sec: renormalization}. 
To investigate the divergent terms in the two-point functions, we use the following expression:
\begin{align}
\label{eq: divergence_Pi}
\Pi_{VV^\prime} (q^2) \approx 
    \Pi_{VV^\prime}(0) + q^2 \Pi_{VV^\prime}^\prime (0), 
\end{align}
where the symbol $\approx$ means that the divergence structures on both sides of the equation are matched. 
This equation is valid for all the two-point functions and is not an approximation.

Using this expression, it is straightforward to see that $M_{V\ast}^2$ ($V=W$, $Z$, and $D$), $e_\ast^2$, and $(\hat{\eta}\hat{\varepsilon})_\ast$ are finite, {\it i.e.}, 
\begin{align}
M_{V\ast}^2 \approx e_\ast^2 \approx (\hat{\eta}\hat{\varepsilon})_\ast \approx 0.
\end{align} 
To investigate the remaining quantities $Z_{V\ast}$, $s_{\theta\ast}^2$, $\xi_\ast$, and $(\eta \varepsilon s_\theta)_\ast$, we use the following decompositions of the two-point functions derived from the definitions of the gauge fields in Sec.~\ref{sec: models}:
\begin{align}
& \Pi_{WW} = \Pi_{\hat{1}\hat{1}} + \Pi_{\hat{2}\hat{2}}, \\[10pt]
& \Pi_{AA} = s_\theta^2 \Pi_{\hat{3}\hat{3}} + c_\theta^2 \Pi_{\hat{B}\hat{B}}
	+ \varepsilon^2 c_\theta^2 \Pi_{\hat{D}\hat{D}}
	+ 2 s_\theta c_\theta \Bigl( \Pi_{\hat{3}\hat{B}} - \varepsilon \Pi_{\hat{3}\hat{D}} \Bigr)
	- 2 \varepsilon c_\theta^2 \Pi_{\hat{B}\hat{D}}, \\[10pt]
& \Pi_{ZZ} = c_\xi^2 c_\theta^2 \Pi_{\hat{3}\hat{3}} + 
	c_\xi^2 s_\theta^2 \Pi_{\hat{B}\hat{B}} 
	+ \biggl(c_\xi \varepsilon s_\theta - \frac{ s_\xi }{ \eta } \biggr)^2 \Pi_{\hat{D}\hat{D}}
	- 2 c_\xi^2 s_\theta c_\theta \Pi_{\hat{3}\hat{B}}
	\nonumber \\
	& \hspace{40pt} + 2 c_\xi c_\theta \biggl(c_\xi \varepsilon s_\theta - \frac{s_\xi}{\eta} \biggr) \Pi_{\hat{3}\hat{D}}
	- 2 c_\xi s_\theta \biggl( c_\xi \varepsilon s_\theta - \frac{ s_\xi }{ \eta } \biggr) \Pi_{\hat{B}\hat{D}}, \\[10pt]
& \Pi_{DD} = s_\xi^2 c_\theta^2 \Pi_{\hat{3}\hat{3}} + 
	s_\xi^2 s_\theta^2 \Pi_{\hat{B}\hat{B}} 
	+ \biggl(s_\xi \varepsilon s_\theta + \frac{ c_\xi }{ \eta } \biggr)^2 \Pi_{\hat{D}\hat{D}}
	- 2 s_\xi^2 s_\theta c_\theta \Pi_{\hat{3}\hat{B}}
	\nonumber \\
	& \hspace{40pt} + 2 s_\xi c_\theta \biggl(s_\xi \varepsilon s_\theta - \frac{c_\xi}{\eta} \biggr) \Pi_{\hat{3}\hat{D}}
	- 2 s_\xi s_\theta \biggl( s_\xi \varepsilon s_\theta + \frac{ c_\xi }{ \eta } \biggr) \Pi_{\hat{B}\hat{D}}, \\[10pt]
& \Pi_{ZD} =  s_\xi c_\xi c_\theta^2 \Pi_{\hat{3}\hat{3}}  
	+ s_\xi c_\xi s_\theta^2 \Pi_{\hat{B}\hat{B}} 
	+ \biggl(c_\xi \varepsilon s_\theta - \frac{ s_\xi }{ \eta } \biggr) 
		\biggl(s_\xi \varepsilon s_\theta + \frac{ c_\xi }{ \eta } \biggr)
		\Pi_{\hat{D}\hat{D}}
	- 2 s_\xi c_\xi s_\theta c_\theta \Pi_{\hat{3}\hat{B}}
	\nonumber \\
	& \hspace{40pt}  + \biggl\{ c_\xi \biggl( s_\xi \varepsilon s_\theta + \frac{ c_\xi }{ \eta } \biggr)
		+ s_\xi \biggl( c_\xi \varepsilon s_\theta - \frac{ s_\xi }{ \eta } \biggr)
		\biggr\} \Bigl( c_\theta \Pi_{\hat{3}\hat{D}} - s_\theta \Pi_{\hat{B}\hat{D}} \Bigr), \\[10pt]
& \Pi_{\tilde{Z}A} = s_\theta c_\theta \Bigl(\Pi_{\hat{3}\hat{3}}
		- \Pi_{\hat{B}\hat{B}} \Bigr)
	- \varepsilon^2 s_\theta c_\theta \Pi_{\hat{D}\hat{D}}
	+ (c_\theta^2 - s_\theta^2) \Bigl( \Pi_{\hat{3}\hat{B}}
		- \varepsilon \Pi_{\hat{3}\hat{D}}\Bigr)
	+ 2 \varepsilon s_\theta c_\theta \Pi_{\hat{B}\hat{D}}, \\[10pt]
& \Pi_{\tilde{D}A} =  c_\xi^2 c_\theta^2 \Pi_{\hat{3}\hat{3}} + 
	c_\xi^2 s_\theta^2 \Pi_{\hat{B}\hat{B}} 
	+ \biggl(c_\xi \varepsilon s_\theta - \frac{ s_\xi }{ \eta } \biggr)^2 \Pi_{\hat{D}\hat{D}}
	- 2 c_\xi^2 s_\theta c_\theta \Pi_{\hat{3}\hat{B}}
	\nonumber \\
	& \hspace{40pt} + 2 c_\xi c_\theta \biggl(c_\xi \varepsilon s_\theta - \frac{s_\xi}{\eta} \biggr) \Pi_{\hat{3}\hat{D}}
	- 2 c_\xi s_\theta \biggl( c_\xi \varepsilon s_\theta - \frac{ s_\xi }{ \eta } \biggr) \Pi_{\hat{B}\hat{D}}, 
\end{align}
where $\hat{a}$ ($a=1$, 2, 3) and $\hat{D}$ represent $\hat{W}^a_\mu$ and $\hat{Z}_D$, respectively. 
These decompositions are valid for any $q^2$.

The isospin symmetry implies that the divergence structure of $\Pi_{\hat{a}\hat{a}}(q^2)$ is the same:
\begin{align}
\Pi_{\hat{1}\hat{1}}(q^2) \approx \Pi_{\hat{2}\hat{2}} (q^2) \approx \Pi_{\hat{3}\hat{3}} (q^2).
\end{align}
In addition, the tracelsssness of $I^3$ leads to 
\begin{align}
\Pi_{\hat{3}\hat{B}}^\prime (0) \approx \Pi_{\hat{3}\hat{D}}^\prime (0) \approx 0. 
\end{align}
Using these facts, we can show 
\begin{align}
& Z_{W\ast} \approx Z_{Z\ast} \approx Z_{D\ast} \approx \xi_\ast \approx 0. 
\end{align}

The right-hand sides of $\Pi_{VV'}$ are now left with possible divergences coming from $s_{\theta\ast}^2$:
\begin{align}
s_{\theta\ast}^2 \approx -\frac{1}{M^2}
	\biggl\{ & \biggl( c_\theta^2 - \frac{ 1 }{ c_\theta^2 } \biggr)\Pi_{\hat{3}\hat{3}}(0)
		+ s_\theta^2 \Pi_{\hat{B}\hat{B}}(0)
		+ \varepsilon^2 s_\theta^2 \Pi_{\hat{D}\hat{D}}(0)
		\nonumber \\
		& - 2 s_\theta c_\theta \Pi_{\hat{3}\hat{B}} (0)
		+ 2 \varepsilon s_\theta c_\theta \Pi_{\hat{3}\hat{D}}(0)
		- 2 \varepsilon s_\theta^2 \Pi_{\hat{B}\hat{D}}(0)
		\biggr\}. 
\end{align}
Using the definition of $A^\mu$, we obtain 
\begin{align}
s_{\theta\ast}^2 \approx 
	-s_\theta \biggl( 1 + \frac{ 1 }{ c_\theta^2 } \biggr) \Pi_{\hat{3}A}(0)
	+ \frac{ s_\theta^2 }{ c_\theta } \Pi_{\hat{B}A}(0)
	- \varepsilon \frac{ s_\theta^2 }{ c_\theta } \Pi_{\hat{D}A}(0)
	= 0,
\end{align}
where we have used the Ward--Takahashi (WT) identity for the last expression. 
Thus, $s_{\theta\ast}^2$ is also finite.\footnote{As mentioned in the main text, we assume that the two-point functions can be made in a gauge-invariant way by using the pinch technique if necessary.}

Finally, we discuss $(\eta \varepsilon s_\theta)_\ast$. 
We first consider the case of the dark $Z$ models using RS-A, where the formulas for the dark photon models are given by taking $\varepsilon_Z \to 0$ and $\delta (\eta \varepsilon_Z)^{\overline{\mathrm{MS}}} \to 0$. 
Using the decompositions and the definitions of $\tilde{B}$, $\tilde{Z}^\mu$ and $\tilde{Z}_D^\mu$, the divergence structure is given by
\begin{align}
(\eta \varepsilon s_\theta)_\ast^{\text{DZ-A}} \approx 
	& \ \frac{ \eta \varepsilon s_\theta }{ M^2 } \biggl\{  
		\Pi_{\tilde{Z}\tilde{Z}} (0) + \frac{ 1 }{ s_\xi c_\xi } \biggl( \frac{ M^2 }{ M_Z^2 - M_D^2 } \biggr) \Pi_{\tilde{Z}\tilde{D}}(0)
		\biggr\}
	\nonumber \\[5pt]
	& + \frac{ \eta s_\theta }{ s_\xi c_\xi } \biggl( \frac{ M^2 \varepsilon_Z^{} }{ M_Z^2 - M_D^2 } \biggr) \Pi_{\tilde{B}\tilde{D}}^\prime(0)
	- \delta (\eta \varepsilon_Z)^{\overline{\mathrm{MS}}}, 
\end{align}
where $\tilde{D}$ represents $\tilde{Z}_D$. 

To proceed with the discussion, we use the one-loop expressions of the two-point functions introduced in Ref.~\cite{Peskin:1990zt}.  Let $J_{3}^\mu(x)$, $J_y^\mu(x)$, and $J_d^\mu(x)$ be the currents associated with the $I_3$, $Y$, and $Q_D$ operators, respectively, and we define the one-loop function $\Pi_{ab}(q^2)$ ($a,b=3$, $y$, $d$) such that
\begin{align}
i\eta^{\mu\nu} \Pi_{ab}(q^2) + (\text{$q^\mu q^\nu$ terms})
    = \int \mathrm{d}^4 x \left< J^\mu_a(x) J^\nu_b(0) \right> e^{-iq\cdot x}
    + \delta \Pi_{ab}^{\mu\nu}, 
\end{align}
where $\delta \Pi_{ab}^{\mu\nu}$ represent the momentum-independent contributions from the corresponding four-point interactions. 
Then, the two-point functions are given by
\begin{align}
& \Pi_{\tilde{Z}\tilde{Z}}(q^2) = g_Z^2 \Bigl\{ 
    c_\theta^4 \Pi_{33}(q^2) - 2 s_\theta^2 c_\theta^2 \Pi_{3y}(q^2) + s_\theta^4 \Pi_{yy}(q^2) \Bigr\}, \\
& \Pi_{\tilde{Z}\tilde{D}}(q^2) = g_Z^2 \Bigl\{ \eta \varepsilon s_\theta c_\theta^2 \Pi_{3y}(q^2) + a \eta c_\theta^2 \Pi_{3d}(q^2) - \eta \varepsilon s_\theta^3 \Pi_{yy}(q^2) - \eta a s_\theta^2 \Pi_{yd}(q^2) \Bigr\}, \\ 
& \Pi_{\tilde{B}\tilde{D}}(q^2) = g_Z^2 \Bigl\{ \eta \varepsilon s_\theta^2 \Pi_{yy}(q^2) + a \eta s_\theta \Pi_{yd}(q^2) \Bigr\},  
\end{align}
where $a = g_D/g_Z$.

Using this representation, the divergence structure is given by
\begin{align}
(\eta \varepsilon s_\theta)_\ast^{\text{DZ-A}} \approx &
    \ \frac{g_Z^2 \eta^2 \varepsilon s_\theta}{s_\xi c_\xi} 
    \biggl( \frac{ \varepsilon_Z^{} }{ M_Z^2 - M_D^2 } \biggr)
    \Bigl( \Pi_{yy}(0) + s_\theta^2 M^2 \Pi_{yy}^\prime(0) + \frac{a s_\theta}{\varepsilon} M^2 \Pi_{yd}^\prime(0) \Bigr)
    \nonumber \\[5pt]
    & - \frac{ g_Z g_D \eta^2 \varepsilon s_\theta  }{ s_\xi c_\xi}
    \frac{ \Pi_{yd}(0)  }{ M_Z^2 - M_D^2 } 
    - \delta (\eta \varepsilon_Z)^{\overline{\mathrm{MS}}},
\end{align}
where we have used the WT identity: $\Pi_{3a}(0) + \Pi_{ya}(0) = 0$ for any $a$. 
Therefore, in the dark $Z$ models, $(\eta \varepsilon s_\theta)_\ast$ can be finite by setting 
\begin{align}
\delta (\eta \varepsilon_Z)^{\overline{\mathrm{MS}}}
=  \mathrm{Div}\biggl[& \frac{g_Z^2 \eta^2 \varepsilon s_\theta}{s_\xi c_\xi} 
    \biggl( \frac{ \varepsilon_Z^{} }{ M_Z^2 - M_D^2 } \biggr)
    \Bigl( \Pi_{yy}(0) + s_\theta^2 M^2 \Pi_{yy}^\prime(0) + \frac{a s_\theta}{\varepsilon} M^2 \Pi_{yd}^\prime(0) \Bigr)
    \nonumber \\[5pt]
    & - \frac{ g_Z g_D \eta^2 \varepsilon s_\theta  }{ s_\xi c_\xi}
    \frac{ \Pi_{yd}(0)  }{ M_Z^2 - M_D^2 } \biggr], 
\end{align}
where $\mathrm{Div}[\cdots]$ means the divergent term in the $\overline{\mathrm{MS}}$ scheme.

In the dark photon models, $\varepsilon_Z = 0$, and there is no counterterm. 
The divergence structure is given by
\begin{align}
(\eta \varepsilon s_\theta)_\ast^{\text{DP}} \approx &
     - g_Z g_D \eta 
    \frac{ \Pi_{yd}(0)  }{ M^2 }. 
\end{align}
One can then prove $\Pi_{yd}(0) \approx 0$ in the dark photon models as follow. 
If $U(1)_Y$ or $U(1)_D$ symmetry is unbroken, $\Pi_{yd}(0) = 0$ due to the WT identity. Thus, the divergence in $\Pi_{yd}(0)$ has to be proportional to the VEV breaking both $U(1)_Y$ and $U(1)_D$ at one-loop level. Since such a VEV would induce the mass mixing $\varepsilon_Z$ in the mass matrix~\cite{Bento:2023flt}, $\Pi_{yd} \approx 0$ in the dark photon models. 
Therefore, $(\eta \varepsilon s_\theta)_\ast$ is also finite in the dark photon models.

\end{document}